\documentclass[journal, twocolumn, 10pt]{IEEEtran}

\usepackage[T1]{fontenc}
\usepackage{lipsum}
\usepackage{blindtext}
\usepackage{multicol}
\usepackage{graphicx}
\usepackage{cite}
\usepackage{amssymb,amsmath}
\usepackage{algorithm}
\usepackage{algpseudocode}
\usepackage{flushend}
\usepackage{multirow}% http://ctan.org/pkg/multirow
    \usepackage{lgrind}
\usepackage{hhline}% http://ctan.org/pkg/hhline
\usepackage{tcolorbox}
\usepackage{graphicx}
\usepackage{cite}
\usepackage{amssymb,amsmath}
\usepackage{algorithm}
\usepackage{color,soul,colortbl}
\usepackage{cleveref}
\usepackage{subfigure}
\usepackage{booktabs}
\usepackage{xcolor}
\usepackage{graphicx}
\usepackage[export]{adjustbox}
\usepackage{caption}
\usepackage{multirow}% http://ctan.org/pkg/multirow
\usepackage{hhline}% http://ctan.org/pkg/hhline
\usepackage{psfrag}

\usepackage{widetext}
\usepackage{flushend}
\usepackage{cuted}
\usepackage{float}
\usepackage{lipsum}

\newcounter{tempEquationCounter}
\newcounter{thisEquationNumber}

\usepackage[compact]{titlesec}         % you need this package
\titlespacing{\section}{0pt}{0pt}{0pt} % this reduces space between (sub)sections to 0pt, for example
\AtBeginDocument{%                     % this will reduce spaces between parts (above and below) of texts within a (sub)section to 0pt, for example - like between an 'eqnarray' and text
  \setlength\abovedisplayskip{0pt}
  \setlength\belowdisplayskip{0pt}}

\usepackage{amsthm}
\newtheorem{theorem}{Theorem}

\theoremstyle{plain}

\theoremstyle{plain}

\theoremstyle{plain}

\providecommand{\lemmaname}{Lemma}
\providecommand{\propositionname}{Proposition}
\providecommand{\theoremname}{Theorem}
\providecommand{\lemmaname}{Lemma}
\providecommand{\propositionname}{Proposition}
\providecommand{\theoremname}{Theorem}
\makeatother
\providecommand{\lemmaname}{Lemma}
\providecommand{\propositionname}{Proposition}
\providecommand{\theoremname}{Theorem}

\definecolor{G}{gray}{0.9}
\definecolor{LC}{rgb}{0.88,1,1}

\usepackage[nolist]{acronym}
\begin{document}
\title{{\LARGE{System Design and Parameter Optimization for Remote Coverage from NOMA-based High-Altitude Platform Stations (HAPS)}}}
\author{{ Sidrah Javed,~\IEEEmembership{Member, IEEE} and Mohamed-Slim Alouini, ~\IEEEmembership{Fellow, IEEE} }
 \thanks{S. Javed  is with Department of Engineering, Durham University, United Kingdom. This research was conducted when she was a post-doctorate researcher at King Abdullah University of Science and Technology (KAUST) and M.-S. Alouini is with CEMSE Division, KAUST, Saudi Arabia.  \\ E-mail: sidrah.javed@durham.ac.uk and slim.alouini@kaust.edu.sa } }

 \maketitle
 \begin{acronym}
\acro{5G}{fifth generation}
\acro{BER}{bit error rate}
\acro{AWGN}{additive white Gaussian noise}
\acro{CDF}{cumulative distribution function}
\acro{CSI}{channel state information}
\acro{FDR}{full-duplex relaying}
\acro{HDR}{half-duplex relaying}
\acro{IC}{interference channel}
\acro{IGS}{improper Gaussian signaling}
\acro{MHDF}{multi-hop decode-and-forward}
\acro{SIMO}{single-input multiple-output}
\acro{MIMO}{multiple-input multiple-output}
\acro{MISO}{multiple-input single-output}
\acro{MRC}{maximum ratio combining}
\acro{PDF}{probability density function}
\acro{PGS}{proper Gaussian signaling}
\acro{RSI}{residual self-interference}
\acro{RV}{random vector}
\acro{r.v.}{random variable}
\acro{HWD}{hardware distortion}
\acro{cHWD}[HWD]{Hardware distortion}
\acro{AS}{asymmetric signaling}
\acro{GS}{geometric shaping}
\acro{PS}{probabilistic shaping}
\acro{HS}{hybrid shaping}
\acro{SISO}{single-input single-output}
\acro{QAM}{quadrature amplitude modulation}
\acro{PAM}{pulse amplitude modulation}
\acro{PSK}{phase shift keying}
\acro{DoF}{degrees of freedom}
\acro{ML}{maximum likelihood}
\acro{MAP}{maximum a posterior}
\acro{SNR}{signal-to-noise ratio}
\acro{SCP}{successive convex programming}
\acro{RF}{radio frequency}
\acro{CEMSE}{Computer, Electrical, and Mathematical Sciences and Engineering}
\acro{KAUST}{King Abdullah University of Science and Technology}
\acro{DM}{distribution matching}
\end{acronym}
\vspace{-10cm}
\begin{abstract}
Stratospheric solar-powered high-altitude platform stations (HAPS) have recently gained immense popularity for their ubiquitous connectivity and resilient operation while providing/catalyzing advanced mobile wireless communication services. They have particularly emerged as the promising alternatives for economical coverage of the remote areas in the world. This makes them suitable candidates to meet the UN Sustainable Development Goals (SDG-2030) for global connectivity. HAPS can provide line-of-sight (LoS) communications to the ground users in its ultra-wide coverage area. We propose to divide these users into multiple user-groups and serve each group by a high-density flexible narrow spotbeam, generated by the phased array antennas mounted on HAPS, to achieve high data rates. We carry out the user association and power allocation in a downlink (DL) non-orthogonal multiple access (NOMA) scheme in each user group. In order to improve the system performance, a sum rate maximization problem is formulated by jointly designing user grouping, user association, beam optimization, and power allocation while guaranteeing the quality-of-service (QoS) for users with limited power budget. We further investigate the outage performance of the users with the proposed approach as compared to the traditional scheme. Our findings reveal the significance of joint design of communication parameters for enhanced system performance, optimum energy utilization, and resource allocation. 
\end{abstract}
%\acresetall
\section{Introduction} 
%%% 6G, Aerial in relation to satellite and terrestrial
%A new paradigm of wireless communication, the sixth-generation (6G) system, 
The upcoming generations of wireless communications envisage an all-coverage network capable of providing ubiquitous and resilient connections for space, air, ground, and underwater. Future wireless communication systems will be driven by hyper-connectivity and human-centricity for persistent, secure and personalized access to digital services and resources \cite{ dang2023human,anjum2022adopting}.
 In this context, non-terrestrial networks will be the indispensable parts of the future communications for seamless extended coverage to the remote areas \cite{araniti2021toward}. Particularly, aerial communications have the capability to not only augment the existing terrestrial and satellite communication infrastructure but also address their numerous limitations. Aerial communication platforms  are favorable for their flexible deployment, ubiquitous coverage, resilient operation, remote coverage, economical 
implementation, and large footprint relative to the terrestrial networks \cite{tariq2020speculative}. Likewise, they are preferred over satellite communications for their quasi-stationary position, low latency, higher area throughput, long-endurance, favorable channel conditions and flexible maintenance/reusability. Thus, they can complement the existing infrastructure by offering ultra-wide coverage for the less privileged, sparsely populated regions or difficult terrains \cite{alsharoa2020improvement}. These characteristics equip them to play a crucial role in bridging the digital divide, by connecting the unconnected parts of the world.

Aerial communications can be realized as Low-Altitude Platform Station (LAPS) or
High-Altitude Platform Station (HAPS). LAPS are the preferred choice for quick deployment for emergency communications in response to some natural disasters. However, they offer limited temporal and spatial coverage based on the battery capacity and flying altitudes, respectively. On the other hand, solar-powered HAPS exhibit green communications, long-endurance, ultra-wide coverage, and self-reliance along with other dividends of aerial communication platforms \cite{karapantazis2005broadband}. This enables solar-powered HAPS as a viable, promising, and versatile candidate for seamless merger and integration with the existing communication infrastructure. The key features of HAPS include the ability to adjust and prioritize capacity or coverage, no special requirements on the user equipment as well as operational spectrum, flexible maintenance and scaling capabilities \cite{shibata2019study}. HAPS can offer a range of applications such as wireless communications, earth observation, remote sensing, surveillance, environmental monitoring, maritime/aviation communications, and atmospheric studies etc. 
Solar-powered HAPS have become viable owing to the recent advances in lightweight composite materials, solar panel efficiency, improved energy conversion/storage, autonomous avionics and antenna beamforming \cite{kurt2021vision,anjum2015terahertz}. HAPS can be realized as a balloon, airship, or aircraft, according to the required power, use cases and cargo capabilities \cite{d2016high}. Interestingly, a solar powered HAPS can maintain its station keeping flight in a circular trajectory to assume a quasi-stationary position for several months based on the favorable atmospheric conditions in the lower stratosphere  \cite{karapantazis2005broadband}.

In the recent times, various contributions have been presented to tackle different challenges in the realization of aerial HAPS communications. For instance, a recent study presents a novel wireless scheme for resource optimization in the integrated satellite-airborne-terrestrial networks for global connectivity \cite{alsharoa2020improvement}. 
Other contributions optimize energy-optimum trajectory planning problem for high altitude long endurance (HALE) aircrafts \cite{FBbolandhemmat2019energy,FBmarriott2020trajectory}. 
Another research proposes system design and performance evaluation of a multi-cell HAPS communication system by employing a steerable adaptive antenna array \cite{Nokiahsieh2020uav}.
Likewise, energy-efficient beamforming for beamspace HAP-NOMA systems is investigated over Rician fading channels \cite{ji2020energy}. Moreover, non-orthogonal multiple access (NOMA) is proposed for effective resource management in MIMO-HAPS over millimeter-wave frequency for backbone network \cite{azzahra2019noma}. Another study proposes deterrence strategy to enable secure communications \cite{anjum2023securitisation}. Massive-MIMO HAPS user grouping and beam forming schemes have been designed using average chordal distances and reduced-dimensional statistical-eigenmodes of the users \cite{lian2019user}.
Likewise, joint user clustering and beamforming for non-orthogonal multiple access networks, termed as JUCAB algorithm, is devised to maximize system utility \cite{kim2020joint}. On the other hand, the joint user association and beamforming for integrated satellite-{HAPS}-ground networks is proposed in \cite{liu2023joint}. 
Other contributions address the minimization of error rates, power consumption and outage probability for HAPS based communication systems  \cite{nauman2017system, yahia2022haps}. However, to the best of authors' knowledge, there is no contribution for jointly addressing the user grouping, user association, beam optimization and multiple access schemes in a HAPS communication system. 

%%%% Aerodynamics, Communications, Energy Budegeting (Harmonize)
The traditional approach of employing a single spot beam to serve all users through orthogonal multiple access is quite inefficient owing to the under utilization of the scarce power and spectrum resources. Therefore, we propose a hybrid approach where the users in the close vicinity are clustered to form user groups in the coverage area. Each user group is then served by a unique spot beam through non-orthogonal multiple access (NOMA). Moreover, the inter-beam interference is circumvented by using time-division multiplex (TDM) between different spot beams. Such implementation requires impeccable link budgeting, sophisticated multiple access techniques, user grouping, user association, beam steering, beamwidth optimization and interference management for quality communication and effective resource management. Therefore, we present a holistic approach and suitable algorithms to tackle the aforementioned challenges while guaranteeing user's QoS, fairness and energy efficiency. The main contributions of this research work are enumerated as:

\begin{itemize}
\item User grouping and number of beams to serve all users in the HAPS coverage area. We employ Geometric Disk Cover (GDC) algorithm to find the optimal number of spot beams and their locations i.e., beam centers and boresight directions within its service area. 
\item User association problem identifies the serving beam for each user. The users of one user-group are served by the same spot beam through NOMA superposition coding. Each user is served by one and only one spot beam for user fairness and interference mitigation. We have proposed greedy algorithm for user association with the objective to maximize the data rates, which is proportional to maximizing the SINR. 
Thus, users join the group with the closest beam center on the horizontal plane rendering higher received SINR.
\item After the user grouping and association, we carry out beamwidth optimization to find the optimal beam locations for the given set of users and minimal beam-width or beam-radius which can serve this user-group. The reduced beamwidth will allow a directive beam with higher power density and antenna gain resulting in the higher received power for each user. The beam optimization 
problem is equivalent to the minimum enclosing circle (MEC) problem which can be solved using the efficient Welzl's algorithm.
\item The resources like power and spectrum are efficiently allocated with the target to maximize performance and minimize interference. The hybrid model uses time-division multiplex to serve different user groups whereas NOMA to serve all the users in the same user group. 
Although phased antenna arrays can produce multiple beams at a time but we have assumed one beam spot at an instance with concentrated power to tackle the incurred small-scale and large-scale fading in the long-distance communication. 
\item Next, we propose power and spectrum efficient NOMA based communications along with the beam-steering phased antenna arrays \cite{Nokiahsieh2020uav}. The aim of NOMA with optimal power allocation is to increase the sum rate while guarantying quality-of-service (QoS), user fairness, and user ordering. The optimal decoding order allows successful successive interference cancellation and perfect decoding at all users. Subsequently, the closed-form solution to the NOMA power allocation problem is presented. 
\item Finally, the outage probability, energy efficiency, spectral efficiency and user fairness of the HAPS communication system are investigated to study the effects of the proposed approaches and algorithms. 
\end{itemize}
The developed novel sequential-iterative algorithm solves the aforementioned problems of user grouping, user association, beam optimization, and resource allocation to maximize the system performance with limited resources. 
%power, spectrum, and time 

%\subsection{Paper Organization and Notation}
%Paragraph 5: Organization
The rest of the report is organized as: Section II describes different aspects of the system under consideration i.e., user grouping, user association, and NOMA. In section III, we study the HAPS communication propagation model to evaluate small-scale and large-scale fading. Moreover, we carry out the link budget analysis after incorporating the phased antenna array directivity gain. Next, we evaluate the system performance in terms of SINR, sum rate, system efficiency, and outage probability analysis in section IV. Moreover, we formulate the optimization problem to maximize system throughput with the given QoS and power constraints. This section V details the optimization, algorithms, and design guidelines of the system parameters to achieve the desired system performance. Later, insightful numerical analysis is carried out in section VI, to quantify the gains obtained with the proposed scheme over the conventional schemes, followed by the comprehensive conclusion in Section~VII.

 In this paper, $x$, $\bf{x}$ and $\bf{X}$ represent the scalar, vector and matrix, respectively. Sometimes, we also use capital symbols for significant scalar variables such as $R$ or $H$. The inverse and absolute value of a scalar variable are respectively denoted as $|x|$ and $x^{-1}$. The symbols $\sum_k$, $\|\bf{w}\|$, $[G]_{\rm dB}$, represent the summation over index $k$, l2-norm of vector $\bf{w}$ and decibel value of gain $G$, respectively. The events are illustrated as $\lbrace R \leq R_{\rm th} \rbrace$ where the probability of this event is given by $ \mathit{Pr}\lbrace R \leq R_{\rm th} \rbrace$. The scalar valued $f(x, y\mid \nu ,\sigma)$ is a function of the independent input variables $x$ and $y$ given $\nu$ and $\sigma$. The exponential and logarithmic function with base $b$ of variable $z$ are denoted by $\exp(z)$ and $\log_b(z)$.  A set is presented as $m \in \{ 1,2,\ldots,M\}$ where $m$ can take any value ranging from $1$ to $M$. Moreover, the notation $\{m \setminus n\}$ indicates any value of $m$ in the described range less $n$.  The notation $R[n]$ is the value of function/variable $R$ at a time instance $n$. Moreover, the notations $\tilde{R}$ and $R^*$ highlight the temporary/transitory variable and the optimal value of the variable in the respective order.

 % The probability of an event $A$ is defined as ${\rm Pr}(A)$. The notations $f_z(z)$ and $f_{z|y}(z|y)$ denote the \ac{PDF}  and conditional PDF of a random variable (r.v.) $z$ given $y$. The operator ${\mathbb{E}}[.]$ denotes the expected value. Considering a r.v. $\Lambda$, the real/in-phase and imaginary/quadrature-phase components of $\Lambda$ are denoted as $\Lambda_{\rm{I}}$ and $\Lambda_{\rm{Q}}$, respectively. Moreover, Additionally, $\mathcal{Z}^{+}$ represents a set of positive integers. $\V{v} = [ v_I  \quad v_Q  ]^{\rm T}$ is the real-composite vector representation of the complex number $v = v_I + i\, v_Q$. 

\section{System Description}\label{SecII}
We consider a typical unmanned solar-powered quasi-stationary HAPS at an altitude $H$ over the desired coverage area with radius $R_{\rm H}$ ranging from $60$km to $400$km as shown in Fig. \ref{fig:HAPS}. The HAPS operate from the stratospheric location (between $18$km to $24$km), pertaining to the lower wind speed and suitable air density in this layer, which is essential for the stable flight operation. HAPS provides communication services to $K$ ground users over the fourth-generation ($4$G) long-term evolution (LTE) or $5$G new radio (NR) air interface via service link and backhaul to the gateway through the feeder link \cite{Nokiahsieh2020uav}. We suppose that the ground users are on the same horizontal plane with coordinates ${\bf u}_k \in \mathbb{R}^2 \quad \forall k$ and their locations are known. We further assume that the phased array antennas are mounted at the bottom of HAPS communication panel to serve the ground users with the high-density flexible narrow spot beams. This section describes the user grouping, user association, NOMA scheme and propagation model with the aim to achieve ubiquitous connectivity with effective resource allocation, minimal power consumption and enhanced system performance.

\subsection{User Grouping}
\begin{figure}[t]
    \centering
   \includegraphics[width=0.9\linewidth]{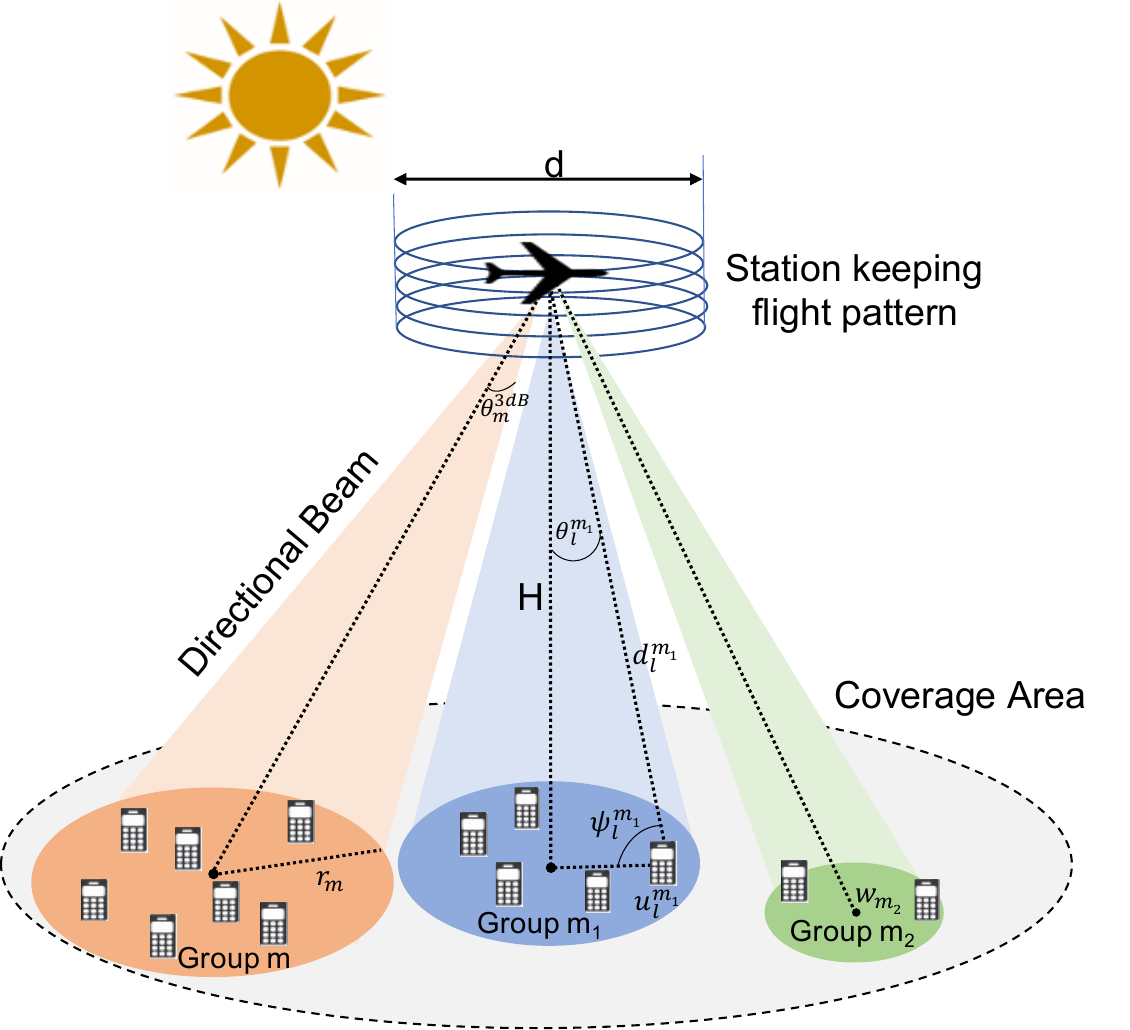}
    \caption{ UAV-based HAPS Communication System  }
    \label{fig:HAPS}
\end{figure}

The ground users in the coverage area are distributed into $M$ groups and each group is served by a high-density narrow spot beam using time-division. This ensures that all users are located in the main lobe of the beam with zero inter-beam interference. The phased array antennas are capable of generating flexible beams with beamwidth $\theta_m^{\rm 3dB} \in \mathbb{R} $, beamradius $r_m \in \mathbb{R}^+$, and beam location\footnote{Beam location indicates the center of the beam with maximum antenna gain} ${\bf w}_m \in \mathbb{R}^2$ for all $m \in \{1,2,\ldots,M \}$ as detailed in Fig. \ref{fig:HAPS}. Interestingly, the number of groups or the number of subsequent beams $M$ is a variable and ranges between $1 \leq M \leq K$, reflecting that there can be atleast one beam to serve all ground users or at max $K$ beams when each user is served by a separate beam. In essence, the choice of $M$ is a trade-off between the number of simultaneously served users and SINR. The higher value of $M$ indicates large number of narrow beams to serve the ground users. This results in increased SINR with high-density power but shorter/delayed time slots. On the other hand, the smaller value of $M$ signifies few but wide beams for coverage. This renders increased transmission time but reduced SINR. The beam coverage radius can be derived from the half-power beamwidth (HPBW) of $m^{\rm th}$ spot beam as:
\begin{equation}\label{eq.BR}
r_m = H \tan \left( \frac{\theta_m^{\rm 3dB}}{2}  \right).
\end{equation}
It is conditioned on $\theta_m^{\rm 3dB} \geq 70 \pi / D$ where $D$ is the diameter of the antenna arrays. Moreover, the beam centers must be chosen in a way that the entire beam coverage area resides within overall coverage area of the HAPS i.e., $\| {\bf w}_m\| + r_m \leq R_{\rm H}$ for effective resource utilization. 

\subsection{User Association}
Assuming $M$ beams to serve $M$ user groups with pre-defined ${\bf w}_m$ and $r_m$, users are associated with groups based on their distances from the center of the beam spots. We define the set of indicators $x_k^m$ to describe users association. The indicator $x_k^m$ is $1$ if user $k$ is associated with group $m$, otherwise 0. It is noteworthy, that each user can only associate with one user group for user fairness and higher efficiency i.e., 
\begin{equation}
\sum_{m=1}^{M} x_k^m = 1, \forall k.
\end{equation}
The beam coverage area is assumed to be circular given by radius $r_m \; \forall \; m$ and all the users in this group are expected to reside in the main lobe of the associated beam to be served simultaneously \cite{liu2020joint}. Therefore, the associated user must fulfill the following constraint:
\begin{equation}
\sum_{m=1}^{M} x_k^m \|{\bf u}_k - {\bf w}_m\| \leq \sum_{m=1}^{M} x_k^m r_m, \quad \forall k. 
\end{equation}
It is noteworthy that a user $l$ may reside within the radius of user-group $j$ while being associated with the user-group $j'$, based on the distance inequality $\|{\bf u}_l - {\bf w}_j\|  \leq \|{\bf u}_l - w_{j'}\|$. Moreover, the apparently overlapping beam radii will not cause any inter-beam interference due to the time-division multiplexing i.e., these user groups are not served simultaneously but at different time intervals.

\subsection{Non-Orthogonal Multiple Access}
Ultra-wide terrestrial coverage of the HAPS can be achieved by splitting the service area into multiple groups, as shown in Fig. \ref{fig:HAPS}. Each group is served by a highly directional beam allowing frequency reuse in the neighboring groups for efficient spectrum allocation. We adopt phased antenna array for beamforming and an array controller which are responsible to create the desired beam and steer it in real time, as detailed in \cite{Nokiahsieh2020uav}. This enables the cell fixation relative to the station-keeping flight pattern for reliable and consistent coverage. 

The network consists of $K$ users distributed as per Poisson Point Process which are partitioned into $M$ groups. In each group coverage, the channel gain is expected to vary with the distance from the center as well as on the azimuth angle. The strongest channel gain is available at the center of the cell along the boresight direction $\theta = 0$. However, as the distance varies and/or the azimuth direction deviates from the boresight, the performance can be degraded due to the increased path loss and lower antenna beam gain. The striking difference in the channel gains of the users in each cell enables us to reap maximum benefits offered by NOMA. Consider the DL-NOMA scenario, where $m^{\rm th}$ group is served by a directional beam with superposition coding as \footnote{The transmitted/received signals, channel gains and allocated powers are function of time. However, the time notation is omitted for brevity.}
\begin{equation}
v_m = \sum_{k=1}^{K}{\sqrt{P_t \alpha_k^m} s_k^m  x_k^m},
\end{equation}
where $P_t $ is the available power budget for transmission after deducting the aerodynamics, electronics, and night-time operational expenses from the available solar power at a given time as quantified in \cite{javed2023interdisciplinary}. 
Moreover, $\alpha_k^m$ and  $s_k^m$ are the fraction of power allocated to and intended information signal for the $k^{\rm th}$ user in $m^{\rm th}$ group, respectively. It is important to highlight that $\sum_{k=1}^{K}{ \alpha_k^m} \leq 1, \; \forall m,$ in order to limit the power division within given budget. Therefore,  using conventional wireless communication model, the received signal at user $l$ in the $m^{\rm th}$ group is given by 
\begin{align}
y_l^m = & \underbrace{h_{lm}^m x_l^m {\sqrt{P_t \alpha_l^m} s_l^m  }}_{\text{Desired Signal}}   +  \underbrace{h_{lm}^m  x_l^m \sum_{k=1 \atop k \neq l}^{K}{\sqrt{P_t \alpha_k^m}  x_k^m s_k^m}}_{\text{IACI}} + w_l^m  x_l^m,
\end{align}
where, $h_{l}^m$ is the channel gain coefficient between the HAPS array panel and $l^{\rm th}$ user in $m^{\rm th}$ cell and $w_l^m$ is the receiver thermal noise modeled as circular symmetric complex Gaussian random variable, i.e., $w_l^m \sim  \mathcal{CN}(0,\sigma_l^2)$. The IACI is the intra-channel interference between users in the same group sharing the same time and frequency resources. In addition, the association parameter $x_k^m$ ensures the incorporation of IACI from the users within the same user-group.

\section{Propagation Model and Link Budget}
\label{SecIII}
The radio signal propagation from HAPS to the UE undergoes free space path loss and multipath fading due to the significant distance between them and obstacles around the UE, respectively\footnote{Note that the HAPS station-keeping flight does not contribute to the fast fading because there are no moving scatters surrounding the aircraft \cite{Nokiahsieh2019propagation}. }. Therefore, the propagation loss of the adopted system is modeled as a combination of both small-scale and large-scale fading. Hence, the channel coefficient $h_{l}^m$ can be expressed as follows:
\begin{equation}\label{eq.Link}
h_{l}^m = \frac{g_{l}^m \sqrt{ G_l^m}}{\sqrt{L(d_l^m)}}, 
\end{equation}
where $g_{l}^m $ is the small scale fading coefficient between the $m^{\rm th}$ transmitting panel and $l^{\rm th}$ user in $m^{\rm th}$ group, $G_l^m$ is the array gain for the link between $m^{\rm th}$ panel and  $l^{\rm th}$ user in its coverage area, and $L(d_l^m)$ is the path loss as a function of $d_l^m$ i.e., the distance between HAPS and $l^{\rm th}$ user in $m^{\rm th}$ group. The computation of these parameters is highlighted in the subsequent sections.

\subsection{Small Scale Multi-path Fading}
The received signal comprises of both the Line-of-Sight (LoS) and Non-LoS (NLoS) components pertaining to the HAPS bore-sight position and independent diffuse multipath reflections from the obstacles. The LoS component is generally deterministic, whereas, the envelope of NLoS component is modeled as a Rayleigh random variable. Hence, the aggregate small-scale multipath fading coefficient $g_{l}^m$ is modeled as a Rician distributed random variable with power distribution function \cite{cuevas2004channel,kanatas2017radio,oladipo2007stratospheric}
\begin{equation}\label{eq.Rician}
f(x\mid \nu ,\sigma_{\rm f })={\frac  {x}{\sigma_{\rm f } ^{2}}}\exp \left({\frac  {-(x^{2}+\nu ^{2})}{2\sigma_{\rm f } ^{2}}}\right)I_{0}\left({\frac  {x\nu }{\sigma_{\rm f } ^{2}}}\right),
\end{equation}
where $I_0$ denotes zeroth-order modified Bessel function of the first kind whose shape parameter $K_s$ is defined by the ratio between the average power of LoS component and the average power associated with NLoS multi-path components i.e., $K_s = {{ {\nu ^{2}}/{2\sigma_{\rm f } ^{2}}}}$.
\subsection{Directivity Gain}

The communication panels are equipped with phased array antenna which are responsible for directional beamforming. The transmitter antenna gain $G_k^m$ of user ${\bf u}_k$ inside $m^{\rm th}$ group depends on the antenna aperture efficiency $\eta$, HPBW of the antenna $\theta_m^{\rm 3dB}$, HAPS altitude $H$, and the distance of the user ${\bf u}_k$ from the center of the beam ${\bf w}_m$ \cite{takahashi2019adaptive}
\begin{equation}\label{eq.Gain1}
\left[ G_k^m \right]_{\rm dB} = \left[ G_{\rm 0}^m  \right]_{\rm dB}  - 12 \frac{G_{\rm 0}^m }{\eta} \left(  \frac{\theta_{{\bf u}_k}}{70 \pi} \right) ^2, 
\end{equation}
where the peak transmitter antenna gain of the $m^{\rm th}$ beam $G_{\rm 0}^m = \eta \left( 70 \pi /  \theta_m^{\rm 3dB} \right)^2 $ and the beam angle (angle of departure) of the user $\theta_{{\bf u}_k}$ is given by
\begin{equation}\label{eq.Gain2}
\theta_{{\bf u}_k} = \tan^{-1} \left( \frac{\|{\bf u}_k - {\bf w}_m\|}{H}    \right).
\end{equation}
 Evidently, the antenna directivity gain reduces while moving away from the boresight position in a horizontal plane.
  
\subsection{Link Budget}
The large scale propagation is characterized as a free space path loss model with the direct distance $d_l^m$ between HAPS and ${u}_l^m$ as $d_l^m = H/\sin \psi_l^m$, where $\psi_l^m$ is the elevation angle of HAPS from  ${u}_l^m$ ranging  ${12 \pi}/{180} \leq \psi_l^m \leq \pi / 2$. Any communication link beyond the minimum elevation angle will lose the LoS path owing to the earth curvature. 
Evidently, the users in the center group enjoy a larger elevation angle, whereas the edge group users are at relatively smaller elevation angles with $0 \leq \psi_l^e \leq \psi_{l'}^c \leq \pi/2$, where $\psi_{l'}^c$ and $\psi_l^e $ are the elevation angles of users $l'$ and $l$ in the center groups and edge groups, respectively. We employ the close-in (CI) path-loss model for the aerial HAPS to compute the received signal path loss $L(d_l^m)$ in dB as \cite{Space2017}
\begin{equation}
[L(H, \psi_l^m)] _{\rm dB} = 10 \log_{10} \frac{16\pi^2 H^2 }{\lambda^2 \sin^2 (\psi_l^m)} + \chi_{\sigma_s}^{\rm CI},
\end{equation}
where $\lambda$ is the wavelength corresponding to the carrier frequency and $\chi_{\sigma_s}^{\rm CI}$ is zero-mean shadow fading Gaussian random variable with standard deviation $\sigma_s$ in dB. The free-space path loss is computed as the ratio between transmit power and received power. The  assumption of isotropic receiver antenna, in the DL-NOMA communication, renders an effective receiver area $\lambda^2 / 4\pi$ and the received signal intensity is given by the inverse-square law. 

\section{Performance Analysis}
The performance of the DL-NOMA with the given user grouping can be analyzed in the form of user's received signal-to-interference noise ratio, sum rate of all users in the groups and outage performance. 
Given the superposition coding at the transmitter for each user group, the received signal undergoes successive interference cancellation to retrieve its own signal based on the known user locations. 

\subsection{Signal-to-Interference Noise Ratio}
Consider the DL-NOMA where $K_m$ is the set of users in the $m^{\rm th}$ group which are ordered as $u_1^m, u_2^m, \ldots, u_{K_m}^m$ depending on their increasing channel strengths. Given this ordered arrangement and successive interference cancellation (SIC) at user ${\bf u}_l^m$, it is capable of decoding all users from $u_1^m$ to $u_{l-1}^m$ and subtracting these from the received signal. Thus, it can decode its own signal from the resultant by considering the interference from $u_{l+1}^m$ to $u_{K_m}^m$ as noise. Therefore, the signal-to-interference noise ratio $\gamma$ at user ${\bf u}_l^m $ is given by
\begin{equation}\label{eq.SINR1}
\gamma_l^m = \frac{|h_{l}^m|^2{{P_t \alpha_l^m}}}{|h_{l}^m|^2 \sum_{k= l+1}^{K_m}{{P_t \alpha_k^m}  } +\sigma^2}.
\end{equation}
whereas the SINR of the user with the strongest channel gain is given by 
\begin{equation}\label{eq.SINR2}
\gamma_{K_m}^m = \frac{|h_{K_m}^m|^2{{P_t \alpha_{K_m}^m}}}{ \sigma^2},
\end{equation}
where the noise power is given as
\begin{equation}
{\sigma^2}({\rm dBm}) = -174 + 10\log_{10}(B) + {\rm NF},
\end{equation}
with NF denoting the noise figure of the receiver \cite{shibata2020system} and $B$ depicting the allocated channel bandwidth to serve $K_m$ users simultaneously. 

\subsection{Sum Rate Analysis}
Assuming perfect receiver channel state information (CSI), we get accurate user-ordering and error-free decoding. Thus, the achievable rate of user ${\bf u}_l^m $ is given by
\begin{equation}
R_l^m =B \log_2 \left[ 1+{\gamma}_l^m  \right].
\end{equation}
conditioned on $R_{j \rightarrow l}^m > {\tilde R}_j^m \; \forall \; j \leq l$, where ${\tilde R}_j^m$ is the targeted data rate of the $j^{\rm th }$ user in the $m^{\rm th }$ usergroup and $R_{j \rightarrow l}^m$ denoted the rate of the $l^{\rm th }$ user to detect $j^{\rm th }$ user's message, $j \leq l$  i.e., 
\begin{equation}
R_{j \rightarrow l}^m  = B \log_2 \left( 1 + \frac{|h_{l}^m|^2{{P_t \alpha_j^m}}}{|h_{l}^m|^2 \sum_{k= j+1}^{K_m}{{P_t \alpha_k^m}  } +\sigma^2} \right) \geq {\tilde R}_j^m. 
\end{equation}
Thus, the sum rate $R$ of all users in $M$ groups can be written as $R = \frac{1}{M}\sum_{m= 1}^{M} R_m$, where the sum rate of all users in $m^{\rm th }$ group is given by $R_m = \sum_{l= 1}^{K_m}{ R_l^m}$ yielding
\begin{equation}
R =\sum_{m = 1}^{M} R_m = B \sum_{m= 1}^{M}\sum_{l= 1}^{K_m} \log_2 \left[ 1+ {  \gamma}_l^m   \right],
\end{equation}
where the received SINR at ${\bf u}_l^m$ i.e., $ {  \gamma}_l^m $ in \eqref{eq.SINR1} can be expressed using \eqref{eq.Link} as
\begin{equation}\label{eq.SINR3}
\gamma_l^m = \frac{ \alpha_l^m}{ \sum_{k= l+1}^{K_m}{{\alpha_k^m}  } + \frac{ L(d_l^m)}{\varrho_m |g_{l}^m|^2 G_l^m } }.
\end{equation}
Likewise $ {  \gamma}_{K_m}^m $ in \eqref{eq.SINR2} can be manifested using \eqref{eq.Link} as
\begin{equation}\label{eq.SINR4}
\gamma_{K_m}^m = \frac{ \alpha_{K_m}^m \varrho_m |g_{K_m}^m|^2 G_{K_m}^m }{ L(d_{K_m}^m)},
\end{equation}
where $\varrho_m$ is the transmit signal-to-noise ratio (SNR) for $m^{\rm th}-$cell i.e., ${P_t}/{\sigma^2}$.
\subsection{System Efficiency}
The system efficiency of the NOMA based HAPS communication system with the proposed user grouping, user association and beam optimization based on the hybrid multiplexing model can be evaluated in terms of energy efficient, spectrum efficiency and user fairness. The energy efficiency of the user $l$ in the $m^{\rm th}$ user group can be investigated using
\begin{equation}
{\rm EE}_l^m = \frac{R_l^m}{\alpha_l^m P_t + P_c  },
\end{equation}
where $P_c$ is the circuit power of the system under consideration. The ${\rm EE}_l^m$ is measured in bits/Joules i.e., a higher value of ${\rm EE}_l^m$ indicates the higher amount of data in bits that can be sent with minimal energy consumption. The overall energy efficiency of the system can be seen as the average energy efficiency of all the users in the coverage area i.e., $ {\rm EE}= (KM)^{-1}\sum_{m=1}^{M} \sum_{l=1}^{Km} {\rm EE}_l^m$.

On the other hand, the spectrum efficiency describes the amount of data transmitted over a given spectrum with minimum transmission errors. Assuming the perfect decoding order and SIC, we can write the spectral efficiency of the user  $l$ in the $m^{\rm th}$ user group as:
\begin{equation}
{\rm SE}_l^m = {R_l^m}/{B_l^m}.
\end{equation}
It is a measure of how efficiently a limited frequency spectrum is utilized to transmit the data by the proposed communication system. It is typically measured in bits/s/Hz. The overall spectral efficient of the NOMA system can be viewed as ${\rm SE} = R/B$ as all the users reap the entire system bandwidth to transmit their data.  
%whereas the same of the OMA system is given by ${\rm SE}_{\rm OMA}=  \sum_{m=1}^{M} R_m/(B/K_m)$ with .]
 In addition, the area spectral efficiency of the $m^{\rm th}$ spot beam in the HAPS communication system can be attained in bits/s/Hz per unit area as:
\begin{equation}
{\rm ASE}_m =  \frac{R_m}{B \pi r_m^2}.
 \end{equation}
Likewise, the area spectral efficiency of the entire system can be viewed as 
\begin{equation}
{\rm ASE} =  \frac{R \tan^2 \left( \psi_{\rm min}  \right)}{B \pi H^2},
 \end{equation}
where the HAPS coverage radius is assumed to be expanding till the LoS viability. 

%The user fairness of a communication system is analyzed to determine whether users or applications are receiving a fair share of system resources. For a given user-group, the user fairness can be quantified using the Jain's fairness index as
% \begin{equation}
% {\mathcal {J}_m}={\frac {(\sum _{i=1}^{K_m}R_{i}^m)^{2}}{K_m \cdot \sum _{i=1}^{K_m}{R_{i}^m}^{2}}}
%  \end{equation}
% The proposed NOMA scheme is particularly designed to elevate user fairness by assigning more transmission power to the users with poor channel conditions and vice versa. 

\subsection{Outage Probability Analysis}
Considering the scenario of unavailable or erroneous CSI, an outage event may happen in NOMA systems. The outage probability can be described as that the $l^{\rm th}$ user is unable to decode its own message or the message of the weaker user $j < l$ in its user cluster/group \cite{ding2014performance}. Thus, the outage probability at the $l^{\rm th}$ user in the $m^{\rm th}$ user group can be written as
\begin{align}
{\rm OP}_l^m= & 1-\mathit{Pr}( \lbrace R_{1 \rightarrow l}^m \geq {\tilde R}_1^m \rbrace \cap  \lbrace R_{2 \rightarrow l}^m \geq{\tilde R}_2^m \rbrace \cap   \ldots \nonumber \\ 
& \ldots \cap  \lbrace R_{l-1 \rightarrow l}^m \geq {\tilde R}_{l-1}^m \rbrace \cap  \lbrace R_{l \rightarrow l}^m \geq {\tilde R}_l^m \rbrace ),
\end{align}
Using the notation $E_{j \rightarrow l}^m$ to denote the event of successful detection of user $j$ message at the $l^{\rm th}$ user and \eqref{eq.Link}, we get
\begin{align}
E_{j \rightarrow l}^m = & \lbrace R_{j \rightarrow l}^m \geq {\tilde R}_j^m \rbrace, \nonumber \\ 
& = \lbrace \frac{|h_{l}^m|^2{{ \alpha_j^m}}}{|h_{l}^m|^2 \sum_{k= j+1}^{K_m}{{ \alpha_k^m}  } +\varrho_m^{-1}} \geq \varphi_j^m \rbrace,
\end{align}
where $\varphi_j^m =   2^{{\tilde R}_j^m / B}-1$. The event $E_{j \rightarrow l}^m$ can be re-written as
\begin{equation}
E_{j \rightarrow l}^m  = \lbrace |h_{l}^m|^2 \geq \frac {\varrho_m^{-1} \varphi_j^m} {{ \alpha_j^m} - \varphi_j^m  \sum_{k= j+1}^{K_m}{{ \alpha_k^m}  }} \rbrace,
\end{equation}
Conditioned on ${{ \alpha_j^m} \geq \varphi_j^m  \sum_{k= j+1}^{K_m}{{ \alpha_k^m}  }}$. Further define
\begin{equation}
\Psi_j^m \triangleq \frac {\varrho_m^{-1} \varphi_j^m} {{ \alpha_j^m} - \varphi_j^m  \sum_{k= j+1}^{K_m}{{ \alpha_k^m}  }}, \qquad \forall j < K_m
\end{equation}
and 
\begin{equation}
\Psi_{K_m}^m \triangleq \frac { \varphi_{K_m}^m} {{ \alpha_{K_m}^m \varrho_m }}. 
\end{equation}
and $\Psi_{\rm max}^{lm}  = \text{max} \lbrace \Psi_{1}^m, \Psi_{2}^m, \ldots , \Psi_{l}^m  \rbrace$. Then, the outage probability can be written as 
\begin{equation}
{\rm OP}_l^m=  1-\mathit{Pr} \left( |h_{l}^m|^2 \geq \Psi_{\rm max}^{lm} \right) = \mathit{Pr} \left( |h_{l}^m|^2 \leq \Psi_{\rm max}^{lm}\right).
\end{equation}
Hence, the outage probability comes out to be the CDF of Rician Squared Distribution $|g_{l}^m|^2$ as
\begin{equation}
{\rm OP}_l^m = \mathit{Pr} \left( |g_{l}^m|^2 \leq \frac{\Psi_{\rm max}^{lm} {L(d_l^m)}}{{G_l^m}}\right).
\end{equation}
Given the users location, user ordering and targeted data rates at an instant, the instantaneous ${\rm OP}_l^m$ can be evaluated, as shown in Appendix A, using the closed form expression as:
\begin{equation}\label{eq.OP}
{\rm OP}_l^m = 1 - \mathcal{Q}_1 \left[ \sqrt{2K_s}, \sqrt{\frac{2 L(d_l^m) \Psi_{\rm max}^{lm} (1+K_s) }{{G_l^m}\Omega}}   \right], 
\end{equation}
where $\Omega$ is the total power from both LoS and NLoS paths, and acts as a scaling factor to the Rician distribution i.e., $\Omega =\nu ^{2}+2\sigma_f^{2}$. On the other hand, the outage probability of a user following OMA just depends on the decoding of its own message. Such that 
\begin{equation}
{\rm OMA}\_{\rm OP}_l^m= \mathit{Pr} \lbrace {\rm OMA}\_R_{l}^m \leq {\tilde R}_l^m \rbrace,
\end{equation}
where 
\begin{equation}
\mathit{Pr} \lbrace {\rm OMA}\_R_{l}^m \leq {\tilde R}_l^m \rbrace =  \mathit{Pr} \lbrace \frac{|h_{l}^m|^2 \rho_m^{\rm OMA}}{K_m} \leq  {\rm OMA}\_\varphi_l^m \rbrace,
\end{equation}
where $ {\rm OMA}\_\varphi_l^m = 2^{{{\tilde R}_l^m K_m}/{B}}-1$.
Hence, the outage probability in an OMA scenario is equivalent to cumulative density function. 
\begin{equation}
{\rm OMA}\_{\rm OP}_l^m =   \mathit{Pr} \left( |g_{l}^m|^2 \leq \frac{\Psi_{\rm OMA}^{lm} {L(d_l^m)}}{{G_l^m}}\right),
\end{equation}
where $\Psi_{\rm OMA}^{lm} = {\rm OMA}\_\varphi_l^m K_m / \rho_m^{\rm OMA}$ with $\rho_m^{\rm OMA} = P_t / \sigma_{\rm OMA}^2$. Note that the orthogonal frequency domain multiple access (OFDMA) based system will allow spectrum segregation among the users and hence the identical noise power/variance $\sigma_{\rm OMA}^2$ for users in $m^{\rm th}$  user-group will be given by
\begin{equation}
{\sigma_{\rm OMA}^2}({\rm dBm}) = -174 + 10\log_{10}(B/K_m) + {\rm NF}.
\end{equation}
Eventually outage probability for OMA system can be written as the following Marcum Q-function
\begin{equation}\label{eq.OMA_OP}
{\rm OMA}\_{\rm OP}_l^m = 1 - \mathcal{Q}_1 \left[ \sqrt{2K_s}, \sqrt{\frac{2 L(d_l^m) \Psi_{\rm OMA}^{lm} (1+K_s) }{{G_l^m}\Omega}}   \right]. 
\end{equation}
We can now formulate the optimization problem to design optimal power allocation in order to maximize the sum rate of all users within the allocated power budget.
%SINR and data rate
\section{Problem Formulation}

This work aims to jointly optimize numerous design parameters with the objective to maximize sum rate of all users in the coverage area of HAPS while guarantying their quality-of-service (QoS), user fairness, and expenses within the available power budget. The optimization problem is targeted at optimizing the following:  
\begin{enumerate}
\item User grouping: $M$ number of user groups to accommodate all user in the coverage area and the central locations for $M$ beam spots i.e., ${\bf w}_m$
\item User association: $x_l^m$ decides the association between the users and the defined groups
\item Beam optimization: Beam width $\theta_m$ or beam radius $r_m$ 
\item NOMA power allocation: Power allocation coefficients for each user in every user group $\alpha_l^m \; \forall \;l,m$
\end{enumerate}
We formulate the design problem for the parameters optimization in a HAPS communication system with the necessary constraints as follows:
\begin{subequations}\label{eq.P1}
\begin{alignat}{3}
\! \! \textbf{P1}\!: &\underset{M,\bf{X,W},\boldsymbol{\theta},\atop \bf {r},\boldsymbol{\alpha}}{\text{maximize }} \!\!
&& \!\!{\sum_{m= 1}^{M}   \sum_{l = 1}^{K_m}  x_l^m  R_l^m }\left( {\alpha_l^m}, \theta_m^{3 {\rm dB}} \right) & \label{eq.p10} \\
&\!\!\!\! \text{s.t.} 
& &  1 \leq M \leq K,& \label{eq.p11} \\ 
& & & x_l^m \in  \lbrace 0,1\rbrace \& \sum_{m=1}^{M} x_k^m = 1, &\forall k  \label{eq.p12} \\
& & &  \sum_{m=1}^{M} x_k^m \!\! \|{\bf u}_k - {\bf w}_m\| \!\! \leq \!\! \sum_{m=1}^{M}\!\! x_k^m r_m,  &\forall k  \label{eq.p14} \\
& & &  \theta_m^{3 {\rm dB}}\! \! \geq \! \!\frac{70 \pi}{ D}  \&  r_m \! \! \geq \! \! \frac {0.443 \lambda H }{ D}, &\forall m \label{eq.p15} \\
& & &    R_{j \rightarrow  l}^m \geq \tilde{R}_j^m, {\rm for} \; j \leq l & \forall m   \label{eq.p17}   \\ 
& & & 0 \leq{  \alpha_l^m } \leq 1,  \& {  \sum_{l = 1}^{K_m}  \alpha_l^m } \leq 1, &  \forall l, m \label{eq.p18} \\
&&& \alpha_1^m \geq \alpha_2^m \ldots \geq \alpha_{K_m}^m,  &\forall K_m, m  \label{eq.p110} 
\end{alignat}
\end{subequations}
where $\bf{X} \in \mathbb{B}^{K{\rm x}M}$ is a Boolean matrix with entries $x_l^m \in  \lbrace 0,1\rbrace, \forall l,m $ where $1 \leq l \leq K$ and $1 \leq m \leq M$ and $\bf{W} \in \mathbb{R}^{M{\rm x}2}$ contains the 2D coordinates of $M$ beam centers. Moreover,  $\boldsymbol{\theta} \in \mathbb{R}^M$ and $\bf {r} \in \mathbb{R}^M$ are the vectors comprising of the 3dB-beamwidth and spot beam radii of $M$ beams. Additionally, $\boldsymbol{\alpha} \in \mathbb{R}^{K{\rm x}M }$ contains the values of the power coefficients for all users in $M$ groups. It is a sparse matrix with non-zero entries only where $x_l^m = 1$. Importantly, the sum of all entries in a row of $\bf{X}$ matrix should be equal to $1$ since any user can only associate to one user group $m$ whereas the sum of all entries in a column of $\boldsymbol{\alpha}$ should be less than or equal to $1$ because the sum of power coefficients of all users within the same user group cannot exceed the allocated power budget. 

The user grouping constraint is given in \eqref{eq.p11} whereas the user association constraints are presented by \eqref{eq.p12}. Any user can only associate to one user group at a given time. Moreover, the constraint \eqref{eq.p14} ensures that the associating user resides within the beam coverage area. The beam constraints \eqref{eq.p15} are essential lower bounds on the 3dB-beamwidth and spot beam radii, which are adjustable by the beamwidth control. Note that a narrower beam than the given bounds is not achievable with the given antenna array dimensions. The target rate constraint guarantees QoS of all users and warrants the accurate decoding of all users with weaker channel gains which is essential for perfect SIC. The last two constraints on the power allocation coefficients \cref{eq.p18,eq.p110} ensure the transmission power expenses with in power budget and optimal user ordering for user fairness. In any given user group, the maximum power is allocated to the user with weaker channel gains and vice versa. 

\begin{algorithm}[!t]
\caption{Geometric Disk Cover Problem}
\label{alg:GDC}
\begin{algorithmic}[1]
\State \textbf{Input}: The coordinates of users $\lbrace {\bf u}_k \rbrace$ in the horizontal plane of the  HAPS coverage area and beam radius $\lbrace r_m \rbrace \; \forall m$.
\State \textbf{Output}: The number $\lbrace M \rbrace$ and the coordinates of the beam centers $\lbrace {\bf w}_m \rbrace$.
\State \textbf{Initialize} $m \gets 1$
\State \textbf{Compute} Distance Matrix ${\bf D}\in \mathbb{R}^{K {\rm x} K}$ containing distance of a user with every other user and evidently zero diagonal entries.
\State \textbf{Define} Boolean matrix $\bar{\bf D}\in \mathbb{B}^{K {\rm x} K}$ with entries $\bar{d}_{ij} = 1$ iff ${d}_{ij} \leq r_m$ otherwise zero.
%\Statex
\While {$ (m \leq K \; || \;   \bar{\bf D} \neq {\bf 0})$}
    \For{$k \gets 1$ to $K$}                    
        \State For each row $k$, find the non-zero entries $\bar{d}_{kj} \neq 0$ and then from $j$ columns, pick one column $l$ with maximum column sum i.e., $l = {\rm argmax}_j \sum_i \bar{d}_{ij}$ 
        \State Mark ${ \bf w}_m$ $\gets$ ${\bf u}_l^m$
        \State Update matrix $\bar{\bf D}$  by nullifying all $j$ columns and $j$ rows which had non-zero entries $\bar{d}_{kj} \neq 0$.
        \State $m \gets m+1$, $k \gets k+j$
    \EndFor
    \EndWhile
    \State \Return $M \gets m$ and ${ \bf w}_m$ indicates the centers of $m$ user groups. 
\end{algorithmic}
\end{algorithm}

The objective of this optimization problem is to maximize the sum rate of all users within the given resources while guarantying QoS and user fairness. However, the problem {\bf P1} is a non-convex mixed integer programming problem in the given optimization variables. Therefore, we divide this problem into sub-problems and solve these sub-problems sequentially as presented in Algorithm \ref{Algo1}. The subproblems are solved for few optimization parameters assuming that rest all design parameters are fixed or given. When the beam coverage radius is fixed, problem {\bf P1} can be converted into the following user-grouping problem:
\begin{subequations}\label{eq.P2}
\begin{alignat}{3}
\!\!\textbf{P1(a)}\! : & \;\underset{M, {\bf W}}{\text{minimize }}
&&{M}  &\label{eq.p20} \\
&\!\!\!\! \text{s.t.} 
& &  1 \leq M \leq K, &\label{eq.p21} \\ 
& & &  x_l^m \in  \lbrace 0,1\rbrace \& \sum_{m=1}^{M} x_k^m = 1, &\forall k \label{eq.p22} \\
& & &  \sum_{m=1}^{M} x_k^m \!\! \|{\bf u}_k - {\bf w}_m\| \!\! \leq \!\! \sum_{m=1}^{M}\!\! x_k^m r_m, \; & \forall k  \label{eq.p24} 
\end{alignat}
\end{subequations}
This sub-problem finds the optimal locations (beam centers ${\bf w}_m$) of the minimal number of beams required to cover the disk of radius $R$ i.e., the coverage area of HAPS communication system. Problem {\bf P1(a)} is a well-known GDC problem which aims to find minimum number of disks of given radius to cover a set of points in a plane. The famous GDC problem is NP-hard highlighting the NP-hardness of {\bf P1}. 
This problem can be solved using Algorithm \ref{alg:GDC}, where the distance and boolean matrices highlight the nearest neighbors with non-zero entries and the user with maximum number of neighbors is marked as the center of user-group ${\bf w}_m$. Next, we eliminate all the users in the coverage neighborhood of ${\bf w}_m$ to find the next beam center. The convergence of the algorithm is guaranteed as it works by eliminating the rows and corresponding columns. The iterations stop when all rows or columns are nullified i.e., all users must reside within the coverage radii of the selected beam centers.

% \begin{lgrind}
%\input Code/meb.m
%  \end{lgrind}

Next subproblem \textbf{P1(b)} solves the user association problem and finds out the association parameters $x_l^m, \; \forall l,m$. 
\begin{subequations}\label{eq.P4}
\begin{alignat}{3}
\!\!\! \textbf{P1(b)} \! :& \underset{{\bf X}}{\text{maximize }}
&& \!\! {\sum_{m= 1}^{M}   \sum_{l = 1}^{K_m}  x_l^m  R_l^m }\left( {\alpha_l^m}, \theta_m^{3 {\rm dB}} \right) & \label{eq.p40} \\
&\!\!\!\! \text{s.t.} 
& &  x_l^m \in  \lbrace 0,1\rbrace, & \forall l,m  \label{eq.p41} \\
& & &  \sum_{m=1}^{M} x_k^m = 1,  & \forall k \label{eq.p42}\\
& & &  \sum_{m=1}^{M} x_k^m \!\! \|{\bf u}_k - {\bf w}_m\| \!\! \leq \!\! \sum_{m=1}^{M}\!\! x_k^m r_m,   & \forall k  \label{eq.p43} 
\end{alignat}
\end{subequations}
Clearly, the users would like to associate with the beams of closest beam centers to receive maximum SINR which will result in the maximum user rate. 
As a result, by greedy algorithm, the indicator variables can be obtained as:
\begin{equation}\label{eq.UA}
x_l^m =
\begin{cases} 
1,  & m = \text{argmin}_m \|{\bf u}_l - {\bf w}_m\|, \\
0,  & otherwise , \\
   \end{cases}
\end{equation}
This \eqref{eq.UA} is evaluated for each user and each user can associate with only one closest beam at a given time. Based on the user grouping and user association from \textbf{P1(a)} and \textbf{P1(b)}, we can carry out beam optimization in the subsequent problem \textbf{P1(c)}.
\begin{subequations}\label{eq.P3}
\begin{alignat}{3}
\textbf{P1(c)}:\quad &\!\!\!\!\underset{{\bf W, r},\boldsymbol{\theta}}{\text{maximize }}
&& {\sum_{m= 1}^{M}   \sum_{l = 1}^{K_m}  x_l^m  R_l^m }\left( {\alpha_l^m}, \theta_m^{3 {\rm dB}} \right) & \label{eq.p30} \\
&\!\!\!\! \text{s.t.} 
&&   r_m \geq \text{max} \lbrace{x_k^m \|{\bf u}_k - {\bf w}_m\|} \rbrace, & \forall k,m \label{eq.p31} \\ 
&&&   \theta_m^{3 {\rm dB}} \geq 70 \pi / D,  & \forall m \label{eq.p32} \\
&&&   r_m \geq 0.443 \lambda H / D, & \forall m  \label{eq.p33} 
\end{alignat}
\end{subequations}
The problem can be solved independently for all user groups. Interestingly, the antenna beam gain \eqref{eq.Gain1} is convex and monotonically decreasing in $r_m$ for a given user group. In addition, $G_l^m$ is directly proportional to $\gamma_l^m$ and eventually $R_l^m$. Moreover, for a fixed HAPS altitude $H$, the beam radius and HPBW are interchangeable as $r_m = H \tan (\theta_m^{\rm 3dB}/2)$.
With this background, we can conclude that maximizing the sum rate or SINR is equivalent to maximizing the antenna beam gain. Notably, the maximization of a function that is convex and continuous, and defined on a set that is convex and compact, attains its maximum at some extreme point of that set \cite{bauer1958minimalplaces}. Hence, the aforementioned problem can be solved by finding the minimum value of beam radius which satisfies the constraints \cref{eq.p31,eq.p33}. This can be achieved in the following two ways: 
\begin{enumerate}
%\item   We can find the optimal beam radius $r_m$ or HPBW $\theta_m^{\rm 3dB}$ by solving the convex problem $\text{maximize}_{\theta_m^{\rm 3dB}} \; \text{min}_l \; G_l^m$ whose solution can be computed by solving $\partial G_l^m / \partial \theta_m^{\rm 3dB} = 0 \; \text{for} \; {\rm max}_l \; \theta_{u_l}$ in the $m^{\rm th}$ user-group, as shown in Appendix \ref{AppendB}: 
%\begin{equation}
%\bar{r}_m = H \tan \sqrt{0.3 \log(10) \theta_{u_l}^2} 
%\end{equation}
%It is the optimal beam radius if it resides within the feasible range, otherwise,
%\begin{equation}
%r_m^* = 
%\begin{cases} 
%\bar{r}_m,     & 0.443 \lambda H /D \leq \bar{r}_m \leq R \\
%0.443 \lambda H /D,    &   0.443 \lambda H /D \geq \bar{r}_m  \\
%R,  & \bar{r}_m \geq R \\
%   \end{cases}
%\end{equation}
%We can further calculate $\theta_m^{\rm 3dB}$ from $r_m^*$ and $w_m$ as the centroid of the locations of all associated users. 
\item The problem is equivalent to solving the MEC problem for a given set of points in the user group thus we can employ the well-known Welzl’s algorithm \cite{welzl2005smallest} to identify the fine-tuned beam locations $W$ with minimum beam radii ${\bf r}$, which can cover the given set of users in a user group. 
\item Another near-optimal solution is to evaluate ${\bf w}_m = K_m^{-1} \sum_{k=1}^K x_k^m {\bf u}_k^m$ and $r_m = {\rm max} \lbrace \| x_k^m{\bf u}_k^m - {\bf w}_m \| \rbrace, \; \forall k$. This simplified closed-form heuristic approach performs close to the optimal solution. 
\end{enumerate}

%\begin{figure}[t]
%    \centering
%   \includegraphics[width=\linewidth, frame]{Code/meb1}
%    \caption{MATLAB program to solve problem \textbf{P1(c)} using Convex Programming. }
%\end{figure}
%
%\begin{figure}[t]
%    \centering
%   \includegraphics[width=\linewidth, frame]{Code/meb2}
%    \caption{MATLAB program to solve problem \textbf{P1(c)} using Welzl's Algorithm. }
%\end{figure}
Given $r_m$, we can evaluate the corresponding HPBW using \eqref{eq.BR} and the process can be repeated independently for each user group. Once the beam optimization problems are solved, we get the optimal user grouping and user association. This enables us to design the power allocation parameters disjointly for each user group based on their distances from the group/beam center and user ordering as shown in \textbf{P1(d)}:

\begin{subequations}\label{eq.P5}
\begin{alignat}{3}
\!\! \textbf{P1(d)}:\! &\! \underset{\boldsymbol{\alpha}}{\text{maximize }} \!\! 
&& \!\!  {\sum_{m= 1}^{M} \!  \sum_{l = 1}^{K_m}  x_l^m  R_l^m }\left( {\alpha_l^m}, \theta_m^{3 {\rm dB}} \right) & \label{eq.p50} \\
&\!\!\!\! \text{s.t.} 
& &  R_{j \rightarrow  l}^m \geq \tilde{R}_j^m, {\rm for} \,  j \leq l & \forall m   \label{eq.p51}   \\ 
& & &  {  \sum\nolimits_{l = 1}^{K_m}  \alpha_l^m } \leq 1,  & \forall m \label{eq.p52} \\
& &&  0 \leq{  \alpha_l^m } \leq 1,& \forall l,m  \label{eq.p53} \\
&&& \alpha_1^m \geq \alpha_2^m \ldots \geq \alpha_{K_m}^m, & \forall K_m, m  \label{eq.p54} 
\end{alignat}
\end{subequations}

%Initialize the current maximum average data rate
%Cmax ← 0; forM=1:Kdo
%Initialize the coordinate of beam centers wm by choosing M users ramdomly;
%repeat
%User grouping
%Calculate {xk,m} according to equation (21); Beam optimization
%Update each wm and rm using the randomized
%algorithm;
%if rm < r0 then rm ← r0;
%end
%if rm > rmax then
%Mark the current M as invalid by CM ← 0;
%End the iteration flag ← 0; end
%until convergence or flag == 0;
%Calculate the current average data rate CM by
%equation (7);
%if CM > Cmax then
%Cmax ←CM;
%Record the current {wm}, {rm} and {xk,m}; end

\begin{algorithm}[!t]
\caption{HAPS Communication Parameters: Optimization }\label{Algo1}
\begin{algorithmic}[1]
\State \textbf{Input}: $\lbrace R \rbrace$, $\lbrace H \rbrace$, $\lbrace K \rbrace$, and $\lbrace {\bf u}_k \rbrace$. 
\State \textbf{Output}: $\lbrace M \rbrace$, $\lbrace {\bf w}_m \rbrace$, $\lbrace r_m \rbrace$, $\lbrace \theta_m \rbrace$, $\lbrace x_l^m \rbrace$, and $\lbrace \alpha_l^m \rbrace \; \forall l,m$.
\State \textbf{Initialize} 
 $i \gets 0$,  ${R}[i-1] \gets R_0$ and  $\epsilon \gets \infty$
 \State \textbf{Select} QoS minimum rate threshold $\Omega_m$ and minimum possible beam radius $r_{\rm min}$
  \State \textbf {Set} tolerance $\delta$,  $r[i] = r_{\rm min}$, and $r_{\rm UB} = R $
  \State \textbf {Choose} $\Delta r$ and identical beam radius $r_m[i] = r[i] \forall m$
\While {$\epsilon \ge \delta$ \& $r_{\rm min} \leq r_m[i] \leq R$} 
\State \textbf{Let}  ${i \gets i+1}$ 
\State \textbf{Update} $r_m[i]= r_m[i-1] + \Delta r$ for all user-groups ensuring sequential increment with every iteration.    
\State \textbf{Determine} $M[i]$ and ${\bf w}_m[i] \; \forall \; m \in \left[  1,M \right]$ using GDC to solve \textbf{P1(a)} given constant $r[i]$. 
\State \textbf{Associate} users by solving \textbf{P1(b)} to evaluate $X[i]$ containing $x_l^m[i]$.
\State \textbf{Optimize} individual beams for each user group to valuate $\tilde{w}_m[i]$, $\tilde{\theta}_m[i]$ and $\tilde{r}_m[i]$ by solving \textbf{P1(c)}.
\State \textbf{Update}  ${w}_m[i] \gets \tilde{w}_m[i]$, $r_m[i] \gets \tilde{r}_m[i]$ and ${\theta}_m[i] \gets \tilde{\theta}_m[i]$. 
\State \textbf{Evaluate} the distance $d_l^m[i]$, elevation angle $\psi_l^m[i]$ and transmit antenna gain $G_l^m[i]$ for each user, in the $m^{\rm th}$ group, from the HAPS station based on ${w}_m[i]$, ${r}_m[i]$ and ${\theta}_m[i]$. 
\State \textbf{Calculate} the Rician channel coefficient using the small scale fading CSI $g_l^m[i]$, pathloss $L(d_l^m[i])$, and beam gain $G_l^m[i]$ for each user $l$ in all $M$ user-groups. 
\State \textbf{Obtain} the available transmit power $P_t [i]$ of a solar powered HAPS at the chosen location on a given date and time of the day using the power estimation algorithms in \cite{javed2023interdisciplinary}. 
\State \textbf{Compute} the power allocation coefficients  $\alpha_l^m[i]$ for each user in the $m^{\rm th}$ user group using the closed form solutions of \textbf{P1(d)} for the given $\Omega_m$ and user ordering.
\State \textbf{Evaluate} the users rate $R_l^m[i]$ using $\alpha_l^m[i]$, channel gains $h_l^m[i]$ to find $\tilde{R}[i]$. 
\State \textbf{Compare} $\tilde{R}[i]$ with $R[i-1]$
\State if $\tilde{R}[i] \geq R[i-1]$: $r_m[i] \gets {\tilde r}_m[i]$ and $R[i] \gets \tilde{R}[i]$
\State if $\tilde{R}[i] \leq R[i-1]$:  $r_m[i] \gets r_m[i-1]$ and ${R}[i] \gets R[i-1]$
\State Update  ${\epsilon \gets  \tilde{R}[i]- {R}[i-1]}$ 
\EndWhile
\State User Grouping Parameters: $M^* \gets M[i]$, ${\bf w}_m^* \gets {\bf w}_m[i]$
\State User Association Parameters: $x_l^{m*} \gets x_l^{m}[i] \; \forall \; l,m $
\State Beam radii: $r_m^* \gets r_m[i]$ 
\State Half-power beam widths: $\theta_m^* \gets \theta_m[i] \; \forall \;m$
\State Power Allocation Parameters: $\alpha_l^{m*} \gets \alpha_l^{m}[i] \; \forall \; l,m$ 
\State  Sum Rate of Users: $R^* \gets R[i]$
\end{algorithmic}
\end{algorithm}

Assuming the same target rate threshold for all users within a user group $m$ i.e., $\tilde{R}_j^m = \Omega_m, \forall j$, the problem \textbf{P1(d)} can be solved in a closed-form as presented in \cite{wang2019user}:
\begin{theorem}
In the considered power allocation problem, the sum rate and minimum power coefficients of users in $m^{\rm th}-$ group, respectively, are given by
\begin{equation}
{R}_m^* = K_m \Omega_m + B \log_2 \left[ 1+  \frac{1 - \sum\nolimits_{k= 1}^{K_m} \hat{\alpha}_l^m }{ A_{K_m}^m  } \right],   
\end{equation}
\begin{equation}\label{eq.PowerAlloc}
\hat{\alpha}_l^m = \left(2^{\Omega_m'}-1 \right) \left( \sum\limits_{k= l+1}^{K_m}{\hat{\alpha}_k^m} +  A_{l}^m  \right),
\end{equation}
where
\begin{equation}
A_{l}^m = \frac{ L(d_l^m)}{\varrho_m |g_{l}^m|^2 G_l^m} \quad \text{and} \quad \Omega_m'= \frac{\Omega_m }{ B}. 
\end{equation}
if the following condition holds
\begin{equation}
\left(  2^ {\Omega_m' }-1 \right)   \left(  \sum\limits_{i= 1}^{K_m}  2^{(i-1)\Omega_m' } A_{i}^m  \right) \leq 1.
\end{equation}
\end{theorem}
 The first term of ${R}_m^* $ is the QoS thresholds of all users in $m^{\rm th}-$ user group whereas the second term is the additional rate of $K_m$ user \footnote{It is important to highlight that the users are ordered with decreasing $A_{l}^m $ i.e., $A_{1}^m \geq A_{2}^m \geq \ldots \geq A_{K_m}^m  $.   } after allocating the remaining power  $1-\sum\nolimits_{k= 1}^{K_m} \hat{\alpha}_l^m$ to it, in order to maximize the sum rate. 
\begin{theorem}
For $\sum\limits_{k= 1}^{K_m} \hat{\alpha}_l^m \geq 1 $, there exists a user $u$ in $1\leq u \leq K_m$ which satisfies the following condition:
\begin{equation}
\begin{cases} 
\left(  2^ {\Omega_m' }-1 \right)   \left(  \sum\limits_{i= u+1}^{K_m}  2^{i-1} A_{i}^m     \right)  \leq 1, &  \\
\left(  2^{ \Omega_m'} -1 \right)   \left(  \sum\limits_{i= u}^{K_m}  2^{i-1} A_{i}^m     \right) \geq 1. & \\
   \end{cases}
\end{equation}
Hence, the maximum achievable sum rate is given by
\begin{equation}
{R}_m^*  = \left(K_m-u\right) \Omega_m + B \log_2 \left[ 1+  \frac{\Delta \alpha}{    1-\Delta \alpha +     A_{u}^m  } \right],   
\end{equation}
where 
\begin{equation}
\Delta \alpha = 1-\left(  2^ {\Omega_m'} -1 \right)   \left(  \sum\limits_{i= u+1}^{K_m}  2^{i-1} A_{i}^m     \right).
\end{equation}
\end{theorem}
The first term of ${R}_m^* $ is the QoS thresholds of users from $u+1$ to $K_m$ and the second term is the rate of user $u$ with power $\Delta \alpha$. It signifies that only $u+1$ to $K_m$ can attain QoS threshold with powers $\hat{\alpha}_{u+1}^m, \hat{\alpha}_{u+2}^m, \ldots, \hat{\alpha}_{K_m}^m$, respectively, using \eqref{eq.PowerAlloc}. However, the users before (and including) $u^{\rm th}$ user cannot achieve their target rates. So, the remaining power $\Delta \alpha$ is allocated to the $u^{\rm th}$ user.

   \begin{table}[t]
\caption{The Adopted System Parameters}\label{tab:Value}
\centering
%\begin{tabular}{p{0.25\linewidth}p{0.25\linewidth}p{0.25\linewidth}}
  \begin{tabular}{||c|c||c|c||}
     \hline
              \hline
{$\xi$}&$2^0$ $14$' $2.04$"E& R &$60$km\\
          \hline
{$\chi$}&$53^0$ $28$' $0.48$"N& $K$ &$100-1000$ \\
          \hline
  SS: {$j_d$}& $2460848$ & $D$ & $1.5$m \\
          \hline
 WS: {$j_d$}&  $2461031$  &$S$ & $143 {\rm m}^2$ \\
          \hline
SS: $\alpha_{\rm ext}$ & $0.465$ & $W$ &$165 {\rm kg}$\\
     \hline
WS: $\alpha_{\rm ext}$ & $0.29$ &$H$&$21$km\\
 \hline
$b$,$AR_w$&$35m,30$& $r_{\rm min}$& $5.4640$km\\    
     \hline
$B$&$200$MHz&$\psi_{\rm min}$ & $12 \pi / 180$ \\
     \hline
$T_n$&$870$&$\delta $, NF&$ 1e-4$, $5$dB\\
    \hline
   $k_B$&$1.3800e-23$ &$N_0$&$-174$dBm \\ 
   \hline
   $ \Upsilon$&$10\%$ &$\sigma_f^2$&$1$ \\ 
        \hline
   $f_c$ &$27.5$GHz &$\Upsilon $&$0.1$\\ 
  \hline
         \hline
  \end{tabular}
\end{table}

 We initialize the parameter optimization algorithm by selecting the HAPS altitude $H$ and its coverage area with radius $R$.  Algorithm \ref{Algo1} begins by incorporating the HAPS geographical location at a given time and date to evaluate the available transmit power  \cite{javed2023interdisciplinary} and the QoS user rate minimum threshold $\Omega_m$. We use the coordinates of users in the horizontal plane to determine their distances $d_l$ and elevation angles $psi_l$ from HAPS. The sum rate is initialized as $R[i-1] = R_0$ with uniform power allocation and $R_0$ being the sum rate of all users in the absence of user grouping and beam optimization i.e., $M=1$, $r_m = R$, and ${\bf w}_m$ is the center of the HAPS circular coverage area. The iterative algorithm assigns the lower and upper bounds on the beam radius in each iteration. It uses the branch and bound algorithm to search the optimal beam radius which maximizes the user's sumrate.  The beam radius is chosen as the mid point in the feasible range $r_{\rm LB} \leq r[i] \leq r_{\rm UB}$ where the initial bounds are selected to be $r_{\rm LB}[i] = r_{min}$ and $r_{\rm UB}[i] = R$. The algorithm solves \textbf{P1(a)} to find the number $M[i]$ and centers of spot beams ${\bf w}_m[i]$ each of radius $r[i]$ which cover all the users in the coverage area. Moreover, the problem \textbf{P1(b)} associates users with the nearest ${\bf w}_m[i]$ and renders $x_l^m[i] \; \forall \; l,m$ at iteration $i$. The recently formed user groups are served with $M$ spot beams which are further optimized using Welzl's algorithm to solve \textbf{P1(c)}. This yields the optimal beam parameters $\tilde{w}_m[i]$, $\tilde{r}_m[i]$, and $\tilde{theta}_m[i]$ for each spot beam. The optimized beams with find-tuned beamwidths render the directive antenna gains $G_l^m[i]$ \cref{eq.Gain1,eq.Gain2}, which is then utilized to evaluate the Rician channel coefficient after incorporating the small and large scale fading. Using the aforementioned parameters, the problem \textbf{P1(d)} is solved to find the power allocation coefficient $\alpha_l^m[i]$  for each user (by applying Theorem 1 and Theorem 2), which maximizes the overall sum rate by guaranteeing the QoS constraints and user fairness. The computed sum rate in iteration $[i]$ i.e., $\tilde{R}[i]$ is then compared with the same of previous iteration $R[i-1]$ to revise the beamradius bounds. Sum that $r_{\rm UB}[i+1] \gets r[i]$ and $R[i] \gets \tilde{R}[i]$ if $\tilde{R}[i] \geq R[i-1]$ and on the other hand  $r_{\rm LB}[i+1] \gets r[i]$ and ${R}[i] \gets R[i-1]$ if $\tilde{R}[i] \leq R[i-1]$. Since the aim is to move in the direction which maximizes the overall sum rate. This iterative algorithm repeats till it meets the stopping criteria i.e., until the increase in sumrate is insignificant $\epsilon \leq \delta$. Eventually, the algorithm furnishes the optimum values of the $M*$, ${\bf w}_m^*$, $r_m^*$, $\theta_m^*$, $x_l^{m*}$, $\alpha_l^{m*}$, and $R*$.

 \begin{figure}[t]
  \centering
\subfigure[Users Distribution - PPP]{\label{fig1:Users}\includegraphics[width=0.53\linewidth,trim={2.5cm 0.5cm 1.25cm 1cm},clip]{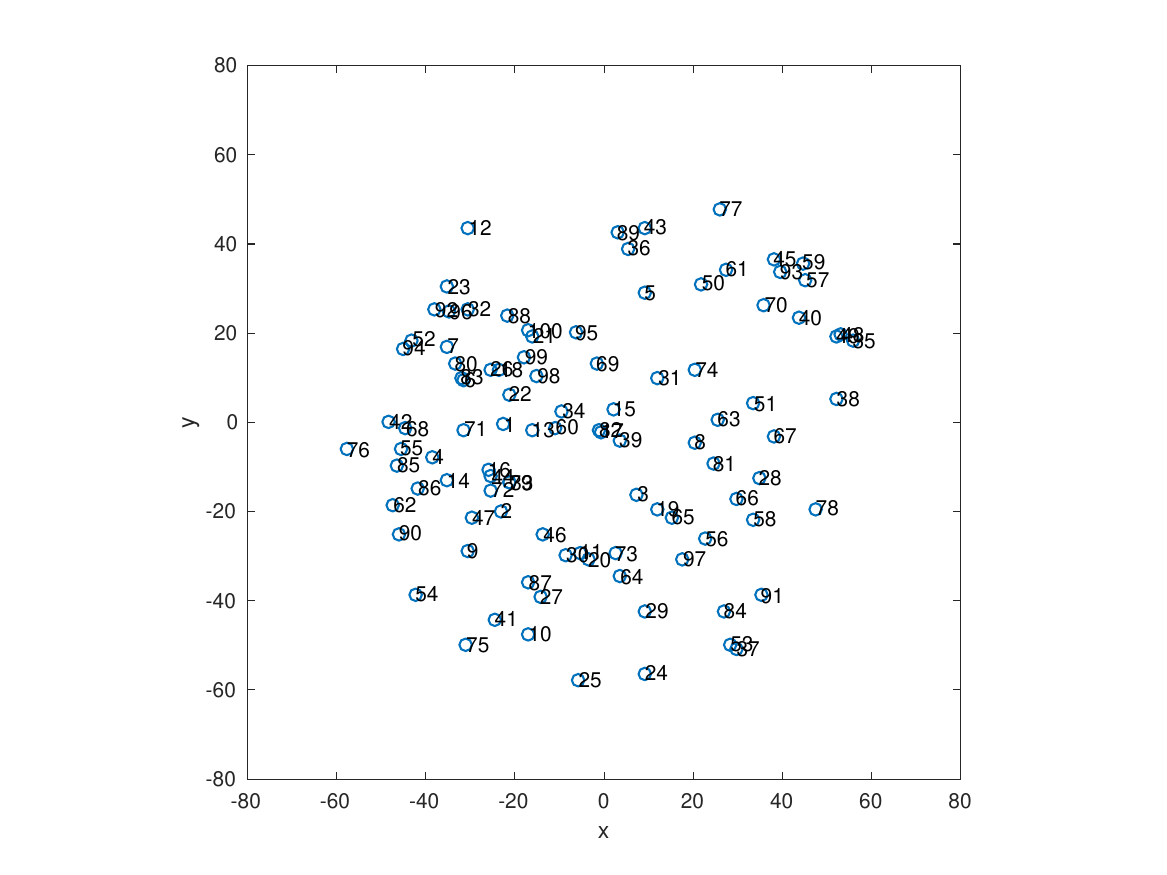}} \hspace*{-2.5em}
\subfigure[User Grouping - GDC]{\label{fig1:GDC}\includegraphics[width=0.53\linewidth,trim={2.5cm 0.5cm 1.25cm 1cm},clip]{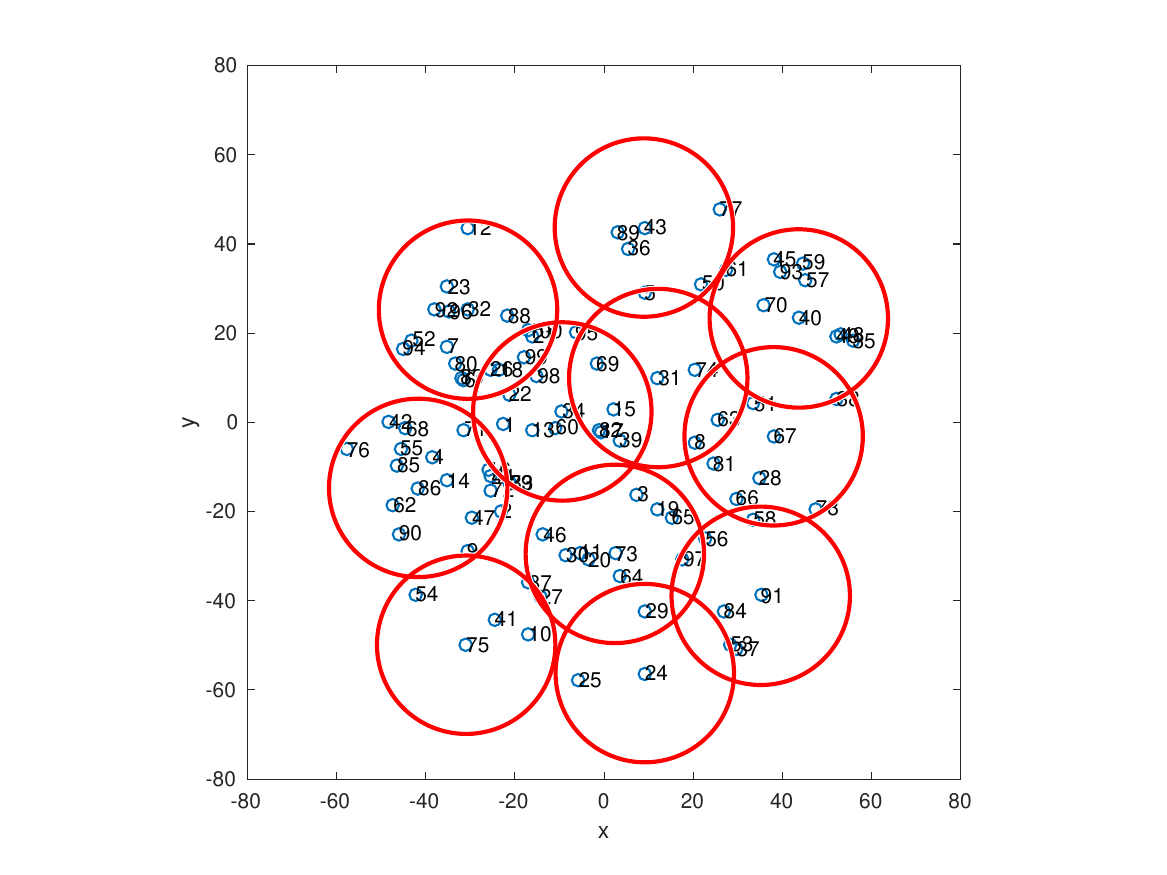}} \\
\subfigure[User Association - GA]{\label{fig1:UA}\includegraphics[width = 0.53\linewidth,trim={2.5cm 0.5cm 1.25cm 1cm},clip]{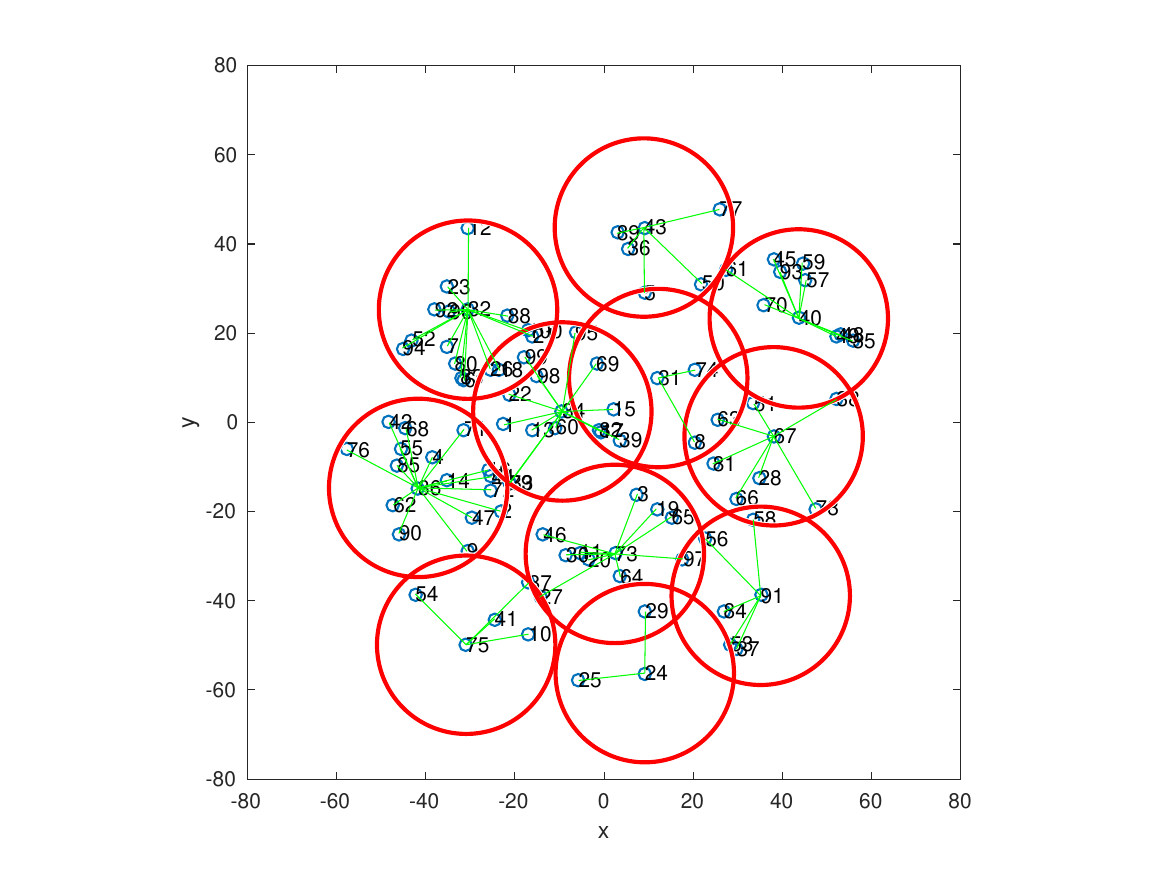}} \hspace*{-2.5em}
\subfigure[Beam Optimization - MEC ]{\label{fig:BO}\includegraphics[width=0.53\linewidth,trim={2.5cm 0.5cm 1.25cm 1cm},clip]{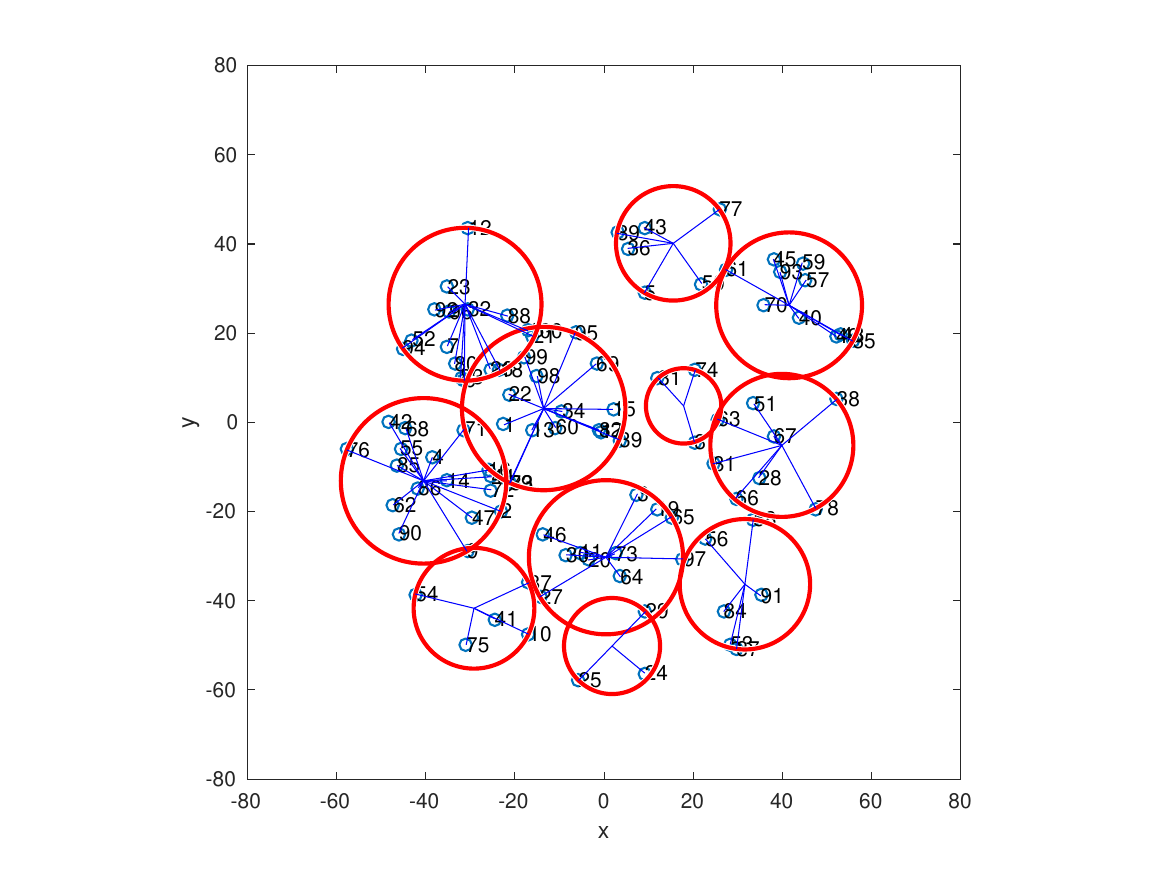}} \\
\caption{User grouping, user association, and beam optimization}
\label{fig:UGA}
\end{figure}

\section{Numerical Results}
The numerical results are evaluated for a HAPS aircraft flying at an altitude $21$km and serving the area with coverage radius of $60$km in Manchester, UK ($2^0$ $14$' $2.04$"E, $53^0$ $28$' $0.48$"N). We have adopted the PHASA-$35$ aircraft model with wingspan $b$ of $35$meters, total weight $W$ (platform and payload mass) $ 165$kg, wing area $S$ $ 143{\rm m}^2$, and maximum achievable altitude $H_{\rm max}$ $21.336$km. We assume $K$ users with Poisson Point Process distribution in the coverage area. We consider $f_c = 27.5$GHz  carrier frequency with $200$MHz channel bandwidth. Moreover, the phased array antenna is assumed to be $90\%$ efficient with diameter $1.5$m. 
The available transmit power is computed for different hours of the day on the winter and summer solstice of 2025 using the solar algorithms \cite{javed2023interdisciplinary}. The boundaries $\psi_{\rm min} =\pi/15$ of the coverage radius is based on the required coverage area as well as the channel characteristics i.e., the LoS link is not available in the region beyond $\psi_{\rm min} =\pi/15$ resulting in a Rayleigh instead of Rician fading channel. The carrier frequency can be adopted from either microwave band or millimeter band depending upon the long-range with limited line losses or higher channel bandwidths preference, respectively. Nonetheless, the choice must be backed by the ITU allocated frequency bands for aerial communications\footnote{ The carrier frequencies of the $2.1$GHz is preferred for seamless merger with the existing terrestrial network \cite{Nokiahsieh2020uav}. It is approved for HAPS base stations offering mobile services according to RR5.388A and ITU Resolution 221(Rec. WRC-07). However, the $27.5$GHz band of FR2 millimeter waves in 5G NR offers a much larger bandwidth with shorter range. Interestingly, the presented system models, performance analysis and optimization framework are valid for any frequency range after incorporating the corresponding path losses in the propagation model.}. The adopted values of numerous simulation parameters are presented in Table \ref{tab:Value} unless specified otherwise.

%%%% Vmin = 12m/s; V_max = 40m/s; Hmin =18; H_o = 18; V_o=40; 
The proposed user grouping, user association, and beam optimization are illustrated in Fig. \ref{fig:UGA}. The Poisson point process distribution of $100$users is depicted in the circular HAPS  coverage area of $60$km centered at the origin $(0,0)$ on a horizontal plane in Fig. \ref{fig1:Users}. The user grouping based on GDC algorithm given $20$km initial beamradius is depicted in Fig. \ref{fig1:GDC}. The algorithm renders the minimum number of $M=11$ beams along with their  optimal locations i.e., beam centers in order to accommodate all users. Next, the user association is carried out based on the greedy algorithm in Fig. \ref{fig1:UA}. The green lines are drawn between the users and the beam centers in a user group to demonstrate their association. Moreover, further beam optimization is exhibited in Fig. \ref{fig:BO} where each spot beam is individually optimized to minimize the beam radius and readjust their centers while serving the same number of users in the assigned user group. Evidently, this reduces the overlapping regions and concentrates the power density in a given beam which maximizes the SINR and consequently the sum rate of the users.

\begin{figure}[t]
    \centering
   \includegraphics[width=\linewidth]{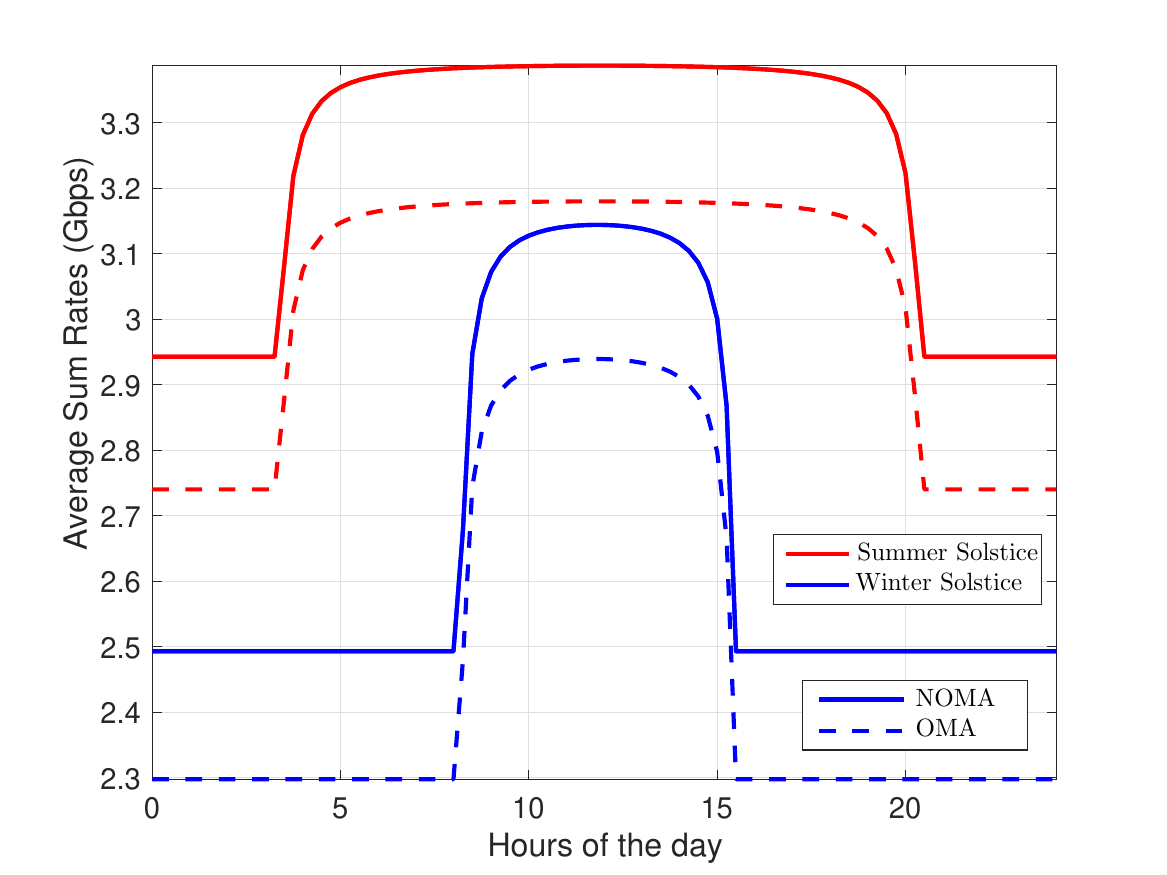}
    \caption{ Average sumrate of the proposed HAPS communication system throughout the day on SS and WS.} \label{fig:RHWS}
\end{figure}

In the next scenario, we present the average sum rate of users in the coverage area of HAPS for different hours of the day on winter solstice (WS) and summer solstice (SS) in 2025. We assume PHASA-$35$ flying over Manchester\footnote{The choice of Manchester is to highlight the sustenance and year long operation of HAPS at such high latitude. Assuming the set \{Alaska, Manchester, Kuala Lumpur, and Jeddah\}, the geographical locations of Kuala Lumpur and Jeddah already enjoy abundant solar irradiance whereas Alaska with limited number of inhabitants cannot sustain year-long operation owing to the very few daylight hours during winters.} in the best and worst case i.e., SS and WS with around $16.75$ and $7$ hours of daylight, respectively. This solar powered HAPS aircraft harvests solar energy and converts it into electrical energy. This energy is then used to fulfill the necessary propulsion, transmission, and accessory expenses while storing the requisite energy for NTO. We propose to utilize all the surplus power during the day for wireless transmission whereas a constant transmission power is preferred for NTO owing to the limited energy during night. The transmission power values are derived from the solar model presented in \cite{javed2023interdisciplinary} after incorporating the inevitable feed line losses. The entire day is divided into the time intervals of $15$ minutes as the solar elevation angle is almost constant in this duration.
The average sum rate with the optimal power allocation in NOMA clearly outperforms the orthogonal multiple access (OMA) counterpart with equal power allocation of all users for both WS and SS, as illustrated in Fig. \ref{fig:RHWS}. Evidently, users can achieve a higher sum rate around noon as compared to the rest of the day pertaining to the higher $P_t$ available at that time. Moreover, the higher sum rates during SS relative to the same in WS is pertaining to the higher available solar power and increased daylight hours.  The percentage increase in the average sum rates at WS and SS is upto $6.97$\% and  $6.5$\% during day and $8.7$\% and $7.08$\% during night, respectively, in the presented scenario.  
\begin{figure}[t]
    \centering
   \includegraphics[width=\linewidth]{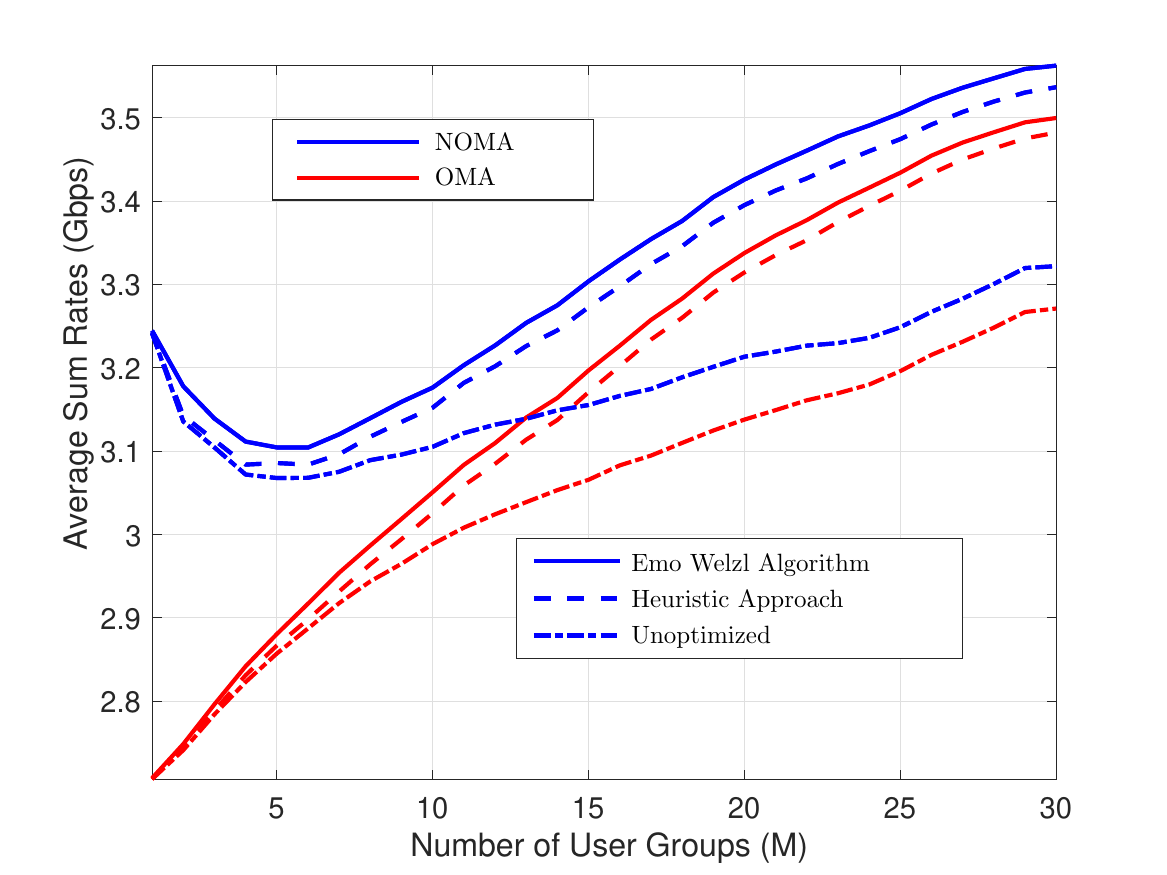}
    \caption{Average sumrate versus number of user-groups employing optimal, sub-optimal or unoptimized spot beams.}\label{fig:RMBO}
\end{figure}

Next, we evaluate the effect of user grouping and beam optimization on the average sum rate for a given set of users in the HAPS coverage area while adopting NOMA and OMA power allocation schemes in Fig. \ref{fig:RMBO}. The average sum rate improvement of $20\%$ can be attained by employing NOMA over OMA in the absence of user grouping and beam optimization i.e., $M=1$. However, the user grouping and beam optimization play significant roles in improving the average sum rate of the users. Interestingly, the sum rate of users with NOMA scheme initially decreases with the distribution in different user groups owing to the TDM. However, the trend changes when the number of user groups is greater than $5$ as the power density of narrow beams dominates the sum rate. Later, the average sum rate is strictly increasing with the increasing number of user groups at the cost of time delays between increasing spot beams. We have employed the optimal Emo Welzl's Algorithm to find the minimum enclosing circles to serve a given set of users and plotted the resultant average sum rate for different number of user groups. In addition, a least-complex heuristic approach is also presented where the beam centers are the mean points of all user locations while the radius is the distance of the farthest user from the beam center in the given set. The heuristic approach performs very close to the optimal beams as compared to the unoptimized beams especially for the higher number of user groups. The data analysis reveals the percentage improvement upto $7.88\%$ and $6.92\%$ by employing Optimal Welzl's algorithm and heuristic approach, respectively, over unoptimized spot beams in case of NOMA. Likewise, we observe the respective percentage improvement upto $7.42\%$ and $6.73\%$ in case of OMA.

\begin{figure}[t]
    \centering
   \includegraphics[width=\linewidth]{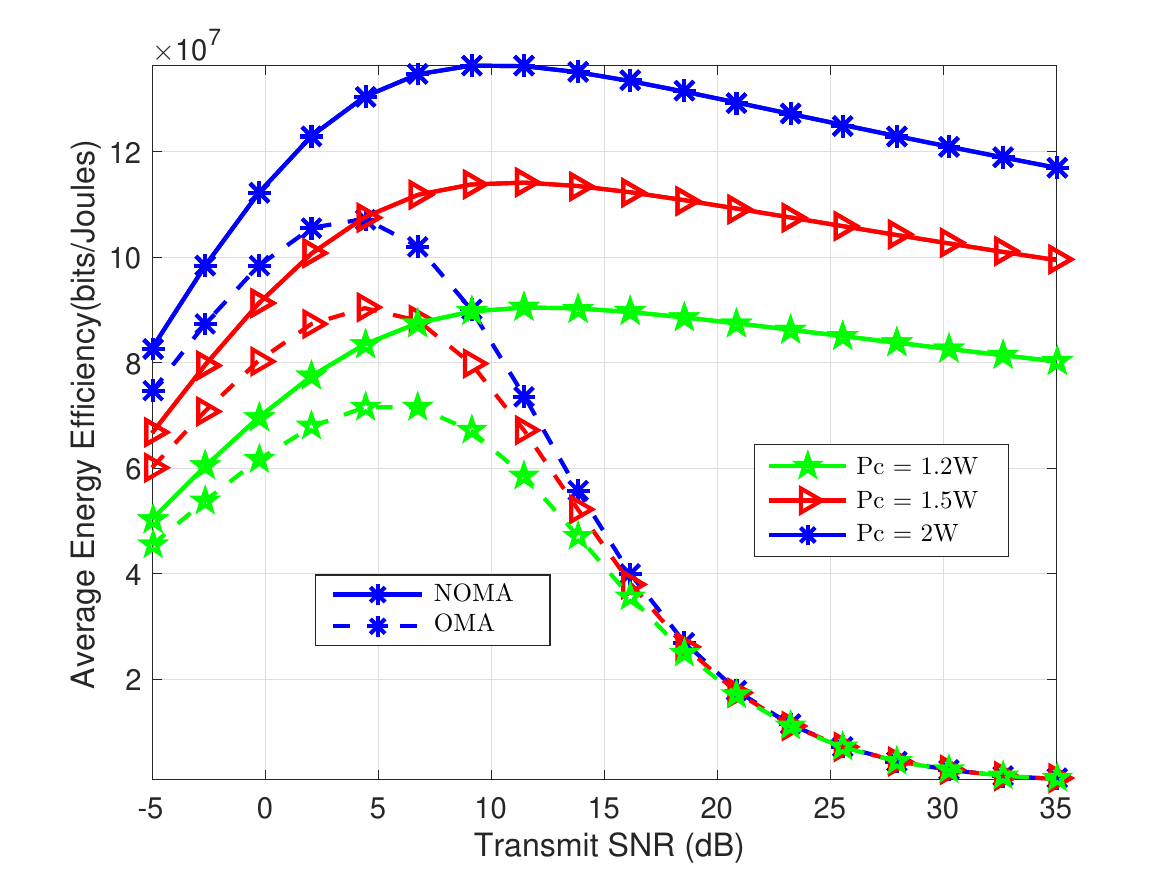}
    \caption{Energy efficiency versus transmit SNR for different circuit power drainage.}\label{fig:EEP}
\end{figure}

We further investigate the average energy efficiency (AEE) of our proposed schemes versus the transmit SNR for different circuit power requirements as demonstrated in Fig. \ref{fig:EEP}. Initially, the AEE increases with the increasing SNR  until it reaches its maximum value at $4dB$ SNR and $10$dB SNR for OMA and NOMA schemes, respectively. Beyond these thresholds, the AEE starts to decrease and eventually saturates. Evidently, the NOMA scheme depicts higher AEE values in the entire range of transmit SNR for three different circuit power cases. We can observe the percentage improvement upto $32\%$, $28.88\%$, and $25\%$ for the $P_c$ values of $1.2$W, $1.5$W, and $2$W, respectively. 
Therefore, we can achieve the maximum energy efficiency at $10$dB transmit SNR for the proposed NOMA scheme in the adopted system.

\begin{figure}[t]
    \centering
   \includegraphics[width=\linewidth]{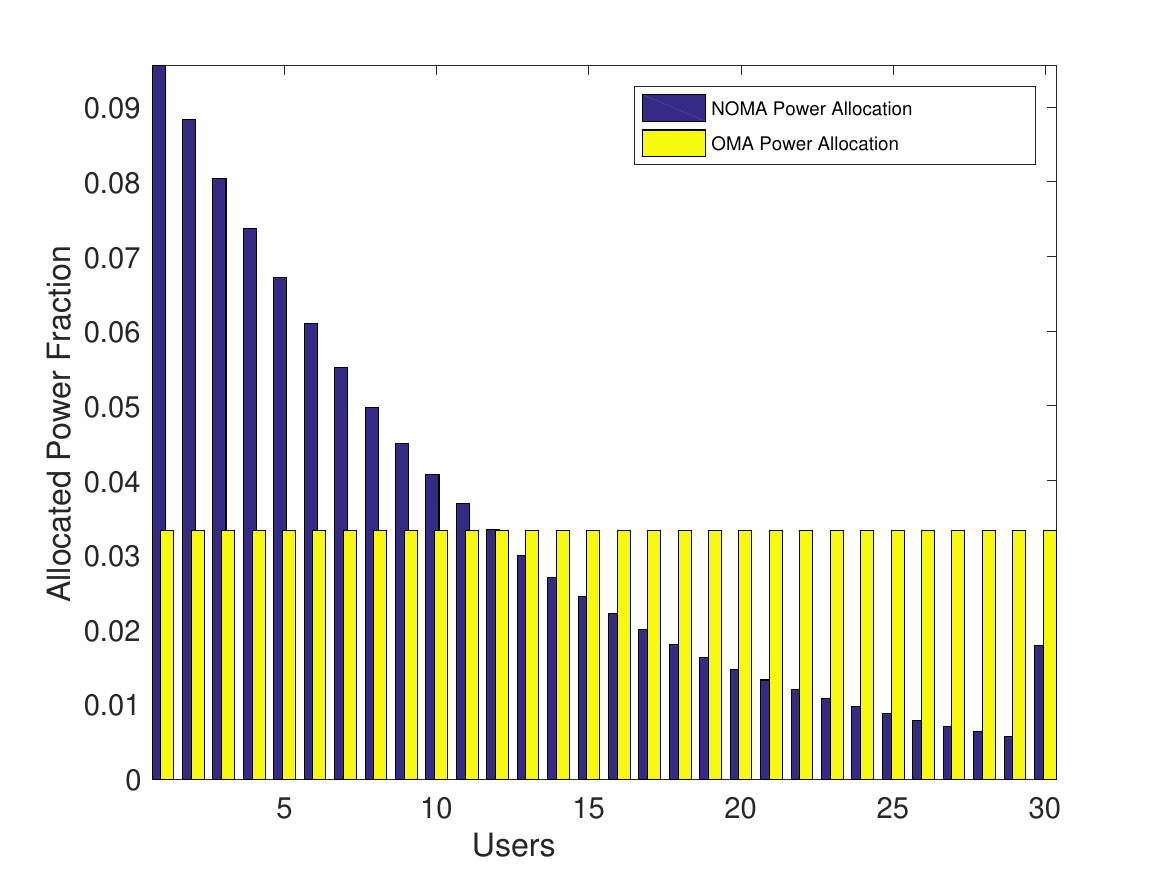}
    \caption{Power allocation amongst users within a usergroup under OMA and NOMA.}\label{fig:PA}
\end{figure}

The bar chart in Fig. \ref{fig:PA} demonstrates the power distribution amongst users in the center cell with NOMA as opposed to the uniform division with OMA. The users with weaker channel strength (edge cell users) are allocated higher fraction of available power budget as compared to the users with stronger channel conditions (center cell users) for a fair distribution, enabling every user in the cell to meet QoS threshold. Once all users are able to meet the target rate, the excessive power is allocated to the strongest user i.e., $u_{K_m} = 30$, which maximizes the overall sum rate as discussed in Theorem 1. On the other hand, for higher QoS threshold only few users with stronger channel conditions can achieve the targets while the weak users experience outage. 

\begin{figure}[t]
    \centering
   \includegraphics[width=\linewidth]{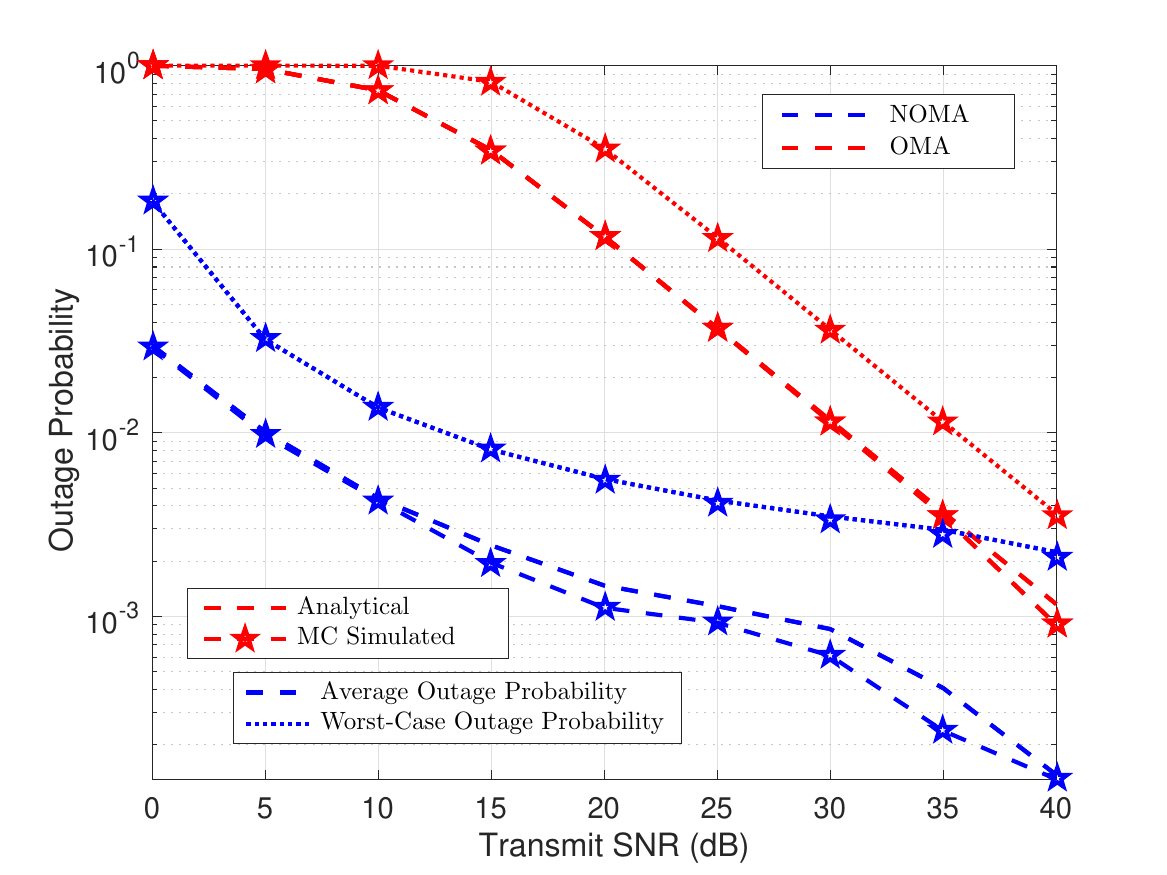}
    \caption{Outage performance of ground users in the proposed system}\label{fig:Pout}
\end{figure}

Eventually, we investigate the outage performance of the proposed HAPS communication system under NOMA and OFDMA for a range of transmit SNR in Fig. \ref{fig:Pout}. In this example, the individual users with achievable data rate below the threshold rate $100$MHz are categorized as users in outage. Expectedly, the number of users in outage decreases with the increase in the available transmit power. Therefore, one can deduce less outage as well as high data rates during the day light hours and summer season. The analysis reveals the lower outage probability of NOMA as opposed to OFDMA for different scenarios. We have presented the average outage probability of all users and worst case outage probability of the edge user for both NOMA and OFDMA. Impressively, the average outage probability for NOMA falls upto $1e-4$ for $40dB$ SNR despite the long-distance communication and excessive path losses whereas OFDMA can barely make it to $1e-3$ at the same SNR level rendering a ten-folds gain with the proposed scheme. On the other hand, the worst-case outage probability presents an error floor meaning that the outage probability cannot be improved with further increase in transmit power. The results in Fig. \ref{fig:Pout} demonstrate high harmony between the presented closed-form analytical expression of outage probability involving Marcum-Q function and the Monte-Carlo simulations. 

\section{Conclusion}
A self-sustaining solar-powered stratospheric HAPS has been proposed to provide aerial communication services to connect the unconnected ground users in its ultra-wide coverage area. The work investigates the effect of employing user grouping, user association and beam optimization on the system performance. We further propose downlink NOMA strategy to superpose information bearing signals of all users in a given user group, which is served by the flexible and steerable beam from the phased array antennas equipped by beam controller.
We have presented different optimization algorithms for optimal and efficient resource management. For instance, geometric disk cover algorithm is proposed for optimal user grouping, greedy algorithm is furnished to solve user association, two different minimum enclosing circle algorithms (i.e., Welzl's algorithm and heuristic approach) are offered for beam optimization and closed-form solution for optimal power allocation amongst users in a given user-group. A holistic and comprehensive Algorithm I is designed to optimize all the aforementioned parameters with the aim to maximize the overall sum rate of the users while guarantying QoS, user fairness, and transmission within the available power budget. 
 The system performance can be analyzed in terms of SINR, average sum rate, link spectral efficiency, mean user spectral efficiency, energy efficiency, user fairness and outage analysis to study the impact of proposed framework. Our findings of the numerical results emphasized the 
joint design of communication parameters to improve the overall system performance. We have achieved upto $20\%$ and { $32\%$} improvement in the average sum rate and energy efficiency, respectively, using NOMA over OMA counterpart while satisfying QoS thresholds of the users. Thus, stratospheric aerial communication platforms can emerge as the promising candidates for global coverage.

\appendices
\section{ Outage Probability Derivation} \label{AppendA}
The probability density function of Rician distributed channel coefficient $|g|$ in \eqref{eq.Rician} can also be written as a function of shape parameter $K_s$ and accumulative power coefficient $\Omega$:  
\begin{equation} \label{eq.Rician1}
f(x \mid K_s ,\Omega)=    {2x \xi } \exp \left( - K_s +  \xi {x^{2} }\right) I_{0}\left(2x \sqrt{ {K_s \xi} } \right),
\end{equation}
where $\xi =  \left( K_s + 1\right) / \Omega $. Using the transformation of variables $|g|^2$, we can get
\begin{equation} \label{eq.Rician2}
f(y \mid K_s ,\Omega) = { \xi } \exp \left( - K_s +  \xi y \right) I_{0}\left(2 \sqrt{ {K_s \xi y} } \right),
\end{equation}
Moreover, the cumulative distribution function of $|g|^2$ can be written in variable $y$, $F(y \mid K_s ,\Omega) = \mathit{Pr} (|g|^2 \leq y)$, as
\begin{equation}
F(y \mid K_s ,\Omega)=1-Q_1 \left(  \sqrt{2 K_s}, \sqrt{2 \xi  y}    \right)
\end{equation}
Given the scaling factors and conditioned on $d_l^m$, the CDF of $|h|^2 = |g|^2 G_l^m  / L(d_l^m)$ can be expressed as
\begin{equation}  \label{eq.CDF}
F(y \mid K_s ,\Omega)=1-Q_1 \left(  \sqrt{2 K_s}, \sqrt{2 \xi  \frac{y L(d_l^m) }{G_l^m} } \right)
\end{equation}
Thus, the outage probability can be evaluated using the CDF in \eqref{eq.CDF} as
\begin{equation}
\text{OP}_l^m = \mathit{Pr} \left( |h_{l}^m|^2 \leq \Psi_{\rm max}^m \right) = F (\Psi_{\rm max}^m \mid K_s ,\Omega) 
\end{equation}
This yields the outage probability expression in \eqref{eq.OP}.

%\section{Proof of Convexity} \label{AppendB}
%We can re-write $G_l^m$ in \eqref{eq.Gain1} as a function of $r_m$ or $\theta_m^{\rm 3dB}$ using the transformation $\theta_m^{\rm 3dB} = 2 \tan^{-1}(r_m /H)  $ as
%\begin{equation}
% G_l^m(r_m)  = 10^{ \log_{10} (\eta ( \frac{70 \pi} { 2 \tan^{-1}(r_m /H)} )^2)  - 1.2 ( \frac{\theta_{{\bf u}_k} }{ 2 \tan^{-1}(r_m /H)})^2}
%\end{equation}
%and
%\begin{equation}
% G_l^m(\theta_m^{\rm 3dB})  = 10^{ \log_{10} (\eta ( \frac{70 \pi} { \theta_m^{\rm 3dB}})^2)  - 1.2 ( \frac{\theta_{{\bf u}_k} }{ \theta_m^{\rm 3dB}})^2}
%\end{equation}
%It is straight-forward to prove that the antenna beam gain $G_l^m$ is convex in $r_m$ or $\theta_m^{\rm 3dB}$ by verifying that $\frac{\partial^2 G_l^m}{\partial {\theta_m^{\rm 3dB}}^2 } \geq 0$, however, the proof is omitted for brevity. Thus, the optimal parameters which 
%
%The partial derivative of $G_l^m$ with respect to $\theta_m^{\rm 3dB}$ is given by
%\begin{align} \label{B3}
%\frac{\partial G_l^m}{\partial \theta_m^{\rm 3dB} } = & 10^{ \log_{10} (\eta ( \frac{70 \pi} { \theta_m^{\rm 3dB}})^2)  - 1.2 ( \frac{\theta_{{\bf u}_k} }{ \theta_m^{\rm 3dB}})^2} \log(10) {\rm x} \nonumber \\ & ( \frac { 12 \theta_{{\bf u}_k}^2 } {5  (\theta_m^{\rm 3dB})^3} -  \frac{2} {\log(10) \theta_m^{\rm 3dB} } )
%\end{align}
%The stationary point can be obtained by equating \eqref{B3} to $0$, which yields
%\begin{equation}
%\bar{\theta}_m^{\rm 3dB} = \sqrt{\frac{6}{5} \log(10) \theta_{{\bf u}_k}^2}
%\end{equation}
%and subsequently
%\begin{equation}
%\bar{r}_m = H \tan \sqrt{0.3 \log(10) \theta_{u_l}^2} 
%\end{equation}

 \bibliographystyle{IEEEtran}
\bibliography{IEEEabrv,refs}

\end{document}